\documentclass[reprint]{revtex4-1}
\usepackage[utf8]{inputenc}
\usepackage{amsmath,graphicx,array,float,amsthm}
\usepackage{mathtools}
\usepackage{xcolor}
\usepackage{romannum}

\newcommand{\figref}[1]{Fig.~\ref{#1}}
\newcommand{\tabref}[1]{Tab.~\ref{#1}}
\renewcommand{\eqref}[1]{Eq.~(\ref{#1})}

\begin{document}

\title{Entanglement-based Quantum Key Distribution in the Daylight and Uplink Satellite Configuration: Proof-of-Principal Demonstration and Feasibility Test}

\author{Kaushik Joarder$^{1,2}$}
\author{Jakub Szlachetka$^{1,2}$}
\author{Piotr Kolenderski$^{1}$}

\affiliation{$^{1}$ Institute of Physics, Faculty of Physics, Astronomy and Informatics, Nicolaus Copernicus University in Torun, ul. Grudziadzka 5, 87-100 Torun, Poland}
\affiliation{$^{2}$ These authors contributed equally to this work.}

\begin{abstract}

Experimental realization of entanglement-based quantum key distribution (QKD) in a daylight uplink (ground-to-satellite) communication channel is highly challenging due to the very low signal-to-noise ratio (SNR) of this configuration and it has yet to be successfully implemented to date. While current research efforts focus on weak coherent pulse (WCP) sources for daylight QKD, we present a proof-of-principle demonstration of uplink-daylight QKD in a free-space-fiber hybrid configuration using polarization-entangled photon pairs generated via a spontaneous parametric down-conversion (SPDC) process. The simulated uplink channel attenuation reaches up to $50~\text{dB}$, equivalent to low-earth-orbit (LEO) distances. Furthermore, the detected noise level (up to a few MHz) in the receiver telescope is representative of daylight conditions. The enhancement of SNR is achieved by implementing rigorous filtering on spatial, spectral, and temporal modes. Our ultra-bright source of entangled photons, characterized by a narrow spectral bandwidth of $0.54~\text{nm}$ FWHM, ensures minimal signal attenuation. We also propose a novel theoretical model of entanglement-based QKD link that fits our experimental data perfectly and is consistent with the prevailing models in the field.  
Our model considers a non-ideal entangled photon source and is applicable to both continuous wave (CW) and pulse-pumped sources. Using this model, we demonstrate that entanglement-based QKD is feasible under uplink daylight conditions in LEO satellites, but only up to a distance of $400~\text{km}$, utilizing current state-of-the-art single-mode fiber (SMF) coupling technology.

\end{abstract}

\maketitle
\vspace{-1.25cm}

\section{Introduction}
\label{sec1}
Global-scale QKD applications necessitate the use of satellite-based optical links. Both terrestrial free-space and optical fiber-based QKD demonstrations have distance limitations of a few hundred kilometers due to the issue of photon absorption and limited line-of-sight \cite{Rarity_2002, RevModPhys.74.145,waks2002security}. On the other hand, the free-space link between the ground and the satellite provides relatively lower photon absorption since the atmosphere is effectively $10-20~\text{km}$ thick, as well as negligible depolarization of the photon state \cite{buttler1998practical}. Both uplink (ground to satellite) and downlink (satellite to ground) QKD channels have been studied in the literature \cite{villoresi2008experimental}. The main advantage of the uplink channel is that the complex single-photon sources, including laser and stabilized interferometers (in case of polarization-entanglement), are in the ground station, and only the detection system containing polarization analyzer and detectors are on the satellite. Hence, the uplink scenario reduces cost and allows miniaturization of the satellite \cite{neumann2018q3sat}. However, one of the downsides of the uplink channel is that the photon loss is higher than the downlink scenario due to the significant beam broadening and wandering caused by the atmospheric turbulence experienced in the initial phase of propagation.

The demonstration of entanglement-based QKD in daylight conditions and the uplink scenario is challenging due to several technical issues. To perform any secure QKD protocol, one must maintain a relatively higher SNR in the detection channel. The signal attenuation is much higher for the uplink scenario than for the downlink case. On the other hand, in daylight conditions, background noise detected by the satellite is nearly six orders higher than in the nighttime \cite{bonato2009feasibility}. Combining these two effects, the SNR value reduces drastically even for the link distance of LEO satellites. Hence, enhancing the SNR by applying tight filtering on the background noisy photons becomes necessary. Three types of filtering are mainly applied to the receiver module: spatial, spectral, and temporal. However, it is also necessary to ensure that these filtering mechanisms do not attenuate the signal photons further. For this reason, in most of the daylight QKD experiments performed so far, WCP created from a laser system with a high pulse repetition rate (few GHz) and/or highly narrow spectral bandwidth (few pm) have been used \cite{shan2006free, liao2017long, ko2018experimental,gong2018free,avesani2021full}. Heralded single photons and entangled photons generated from quantum processes such us SPDC and four-wave mixing usually suffer from relatively lower brightness (few MHz) and higher bandwidth ($>1$ nm). This is mainly the reason for limited experimental attempts \cite{peloso2009daylight} to perform entanglement-based QKD in the daylight scenario, even in the terrestrial free-space channel. Quantum-dot-based entangled photons, although requiring complex experimental setup and with lower collection efficiency compared to the SPDC sources, have also been promising in this regard \cite{basset2023daylight}. 

Selection of the operating wavelength is also an important step. Wavelength of $785~\text{nm}$ seems to be a suitable choice as it falls in the near-infrared band, which is advantageous in the free-space communication considering geometric coupling, spatial filtering, and diffraction losses \cite{lanning2021quantum} while availing low-noise and highly efficient space-qualified commercial single-photon detection technology easily adaptable to a small-scale satellite.

Recent experimental efforts \cite{szlachetka2023ultrabright,steinlechner2012high} to realize stable SPDC-based entangled photon-pair sources with high brightness, entanglement fidelity, and reduced bandwidth have opened up possibilities to perform entanglement-based QKD in daylight and uplink scenarios. Hence, verifying and analyzing the performance of such QKD demonstrations in a controlled experimental environment is necessary. 

In this manuscript, we first describe a novel theoretical model of entanglement-based QKD link for any joint polarization basis measurement on Alice's and Bob's sides. This model considers all realistic experimental imperfections such as realistic entangled photon source, asymmetric detection efficiency, polarization-sensitive losses, dead-time of detectors as well as transmission channel parameters such as background noise, and channel attenuation. This model is suitable for both CW and pulse-pumped sources, providing a more intuitive and comparative understanding of both processes. We extend the model for realistic QKD implementations for both passive and active choice of random basis selection. We also consider the effect of multi-photon pair generation. 

In the experimental part, we distribute entangled photons between two telescopes and measure the polarization visibility while the atmospheric turbulence and daylight noises are simulated inside the free-space channel. We simulate the uplink channel attenuation for varying link distances by attenuating the signal photon counts (from $20$ dB to $50$ dB). Background noise level at the satellite is simulated using a light source emitting unpolarized photons (up to a few MHz rate) with the same central wavelength as the signal photons. Finally, we measure the secret key rate (SKR) and quantum bit error rate (QBER) for a hybrid channel (free-space and fiber) QKD demonstration.

\section{Theory}
\label{sec2}
In this section, we provide a comprehensive theoretical model of the optical link used for entanglement-based QKD between an optical ground station and a LEO satellite. Readers who are more interested in the experimental aspects can skip to Sec. \ref{sec3}.

Numerous excellent theoretical research works are available in the literature \cite{bonato2009feasibility, tomaello2011link, aspelmeyer2003long, er2005background, neumann2018q3sat} on link budget analysis for uplink and downlink scenarios, covering a range of protocols from prepare-and-measure-based to entanglement-based QKD. Theoretical models \cite{ma2007quantum, neumann2021model, takesue2010effects, holloway2013optimal} for SPDC-based entangled photon-pair sources, including multi-photon pair events, have been explored for QKD frameworks. However, as noted in Ref. \cite{neumann2021model}, two separate theoretical models must be considered for CW-pumped and pulse-pumped sources, despite both using the same underlying SPDC process. Therefore, in this section, we propose an alternative model that is equally applicable to both CW and pulsed pumping conditions. This model holds importance from an experimental standpoint as it allows for the consideration of an experimentally reconstructed density matrix of the entangled photon-pair to estimate the SKR and QBER. This approach also aids in distinguishing the noise contribution from a non-ideal entangled-photon source and a noisy communication channel.

\subsection{General theoretical model
for both CW and pulse-pumped entangled photon source}
\label{sec:General theoretical model}
\begin{figure}[h!]
\centering
\includegraphics[height=5cm]{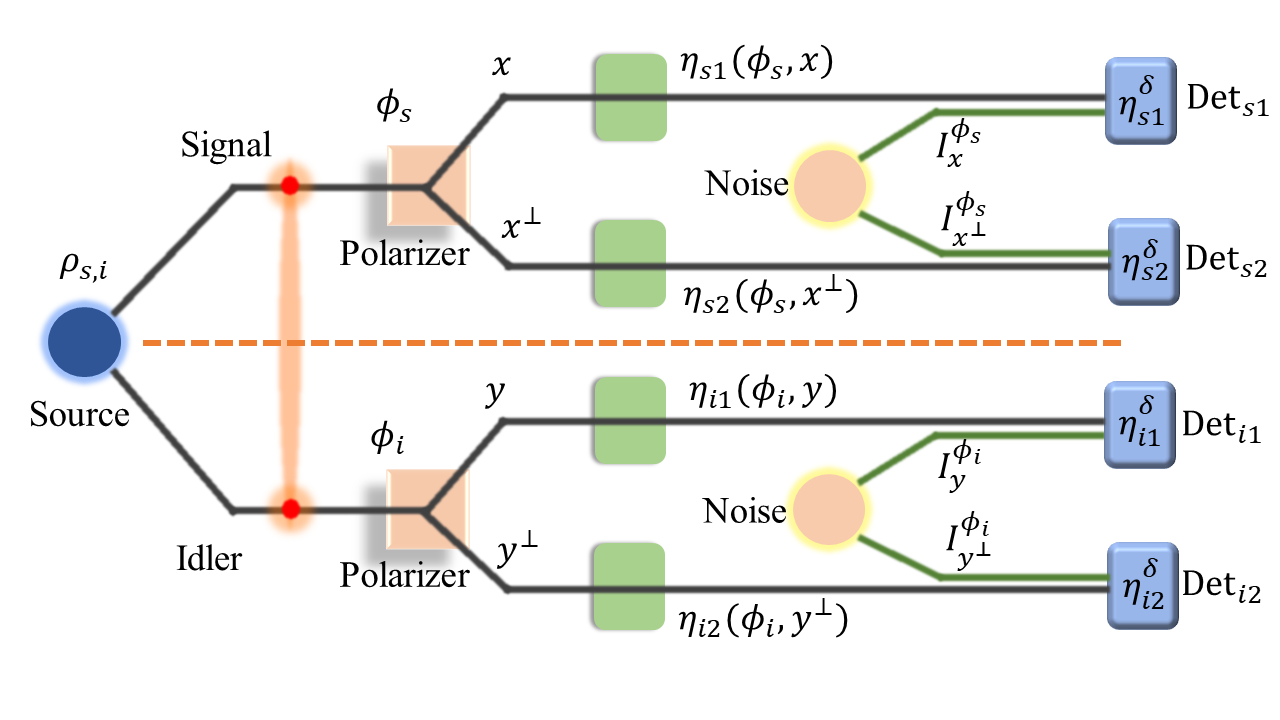}
\caption{Conceptual model depicting the distribution of polarization-entangled photon pairs between two parties. Measurements in the polarization basis are carried out using a polarizer module and single-photon detectors, $\text{Det}_{s(i)}$. The total channel attenuation and background noise (including detector dark counts) are denoted as $\eta_{s(i)}$ and ${I_{x(y)}^{\phi_{s(i)}}}$, respectively. The parameter $\eta^\delta_{s(i)}$ accounts for photon loss due to the detector dead time.}
\label{fig:Sagnac_theory_model}
\end{figure}

A simplified model is presented in \figref{fig:Sagnac_theory_model} to measure the polarization visibility of a polarization-entangled photon-pair source after it passes through a noisy and lossy medium. The performance of the source can be represented by the density matrix of the signal-idler photon-pair, denoted as $\rho_{s,i}$. In both the signal and idler arms, a loss-less polarization splitter, referred to as a polarizer, directs the photon in one of the two orthogonal projections (${x,x^\bot}$ for signal and ${y,y^\bot}$ for idler) for any chosen polarization basis $\phi_{s(i)}\in [-\pi/2,\pi/2]$. The total channel efficiency, including the detector, is denoted as $\eta_{s(i)}$, which depends on the chosen polarization projection on the photon as well as the asymmetric quantum efficiencies of different detectors. Additional noise contribution for each choice of polarization is denoted as ${\mathcal{I}}_{x(y)}^{\phi_{s(i)}}$. Finally, the loss contribution from the dead-time $\delta_{s(i)}$ of each detector $Det_{s(i)}$ is expressed in the efficiency parameter $\eta_{s(i)}^{\delta}$. It is important to note that although the ordering of the optical components is slightly different in the real experimental setup, this simplistic model captures the equivalent effect in a realistic setup.

First, we write the expression for the simplistic case of entanglement distribution. In this case, we need at least one detector in both the signal and the idler arm. The polarizer setup is adjusted to measure any projection $x(y)$ for a chosen basis $\phi_{s(i)}$. For now, we assume that the rate of generating multiple photon pairs is insignificant. Later, we will delve into more realistic scenarios for QKD that involve two and four detectors per channel, as well as consider the contribution of multiple photon pairs.

We measure the coincidence count rate $\mathcal{C}_{x,y}^{\phi_s,\phi_i}$ for any joint projection $(x,y)$ and a given polarization basis $\phi_{s(i)}$ for the signal (idler). $x\in\{0,1\}$, $y\in\{0,1\}$ represents the two orthogonal projections in the $\phi_s$ and $\phi_i$ basis, respectively. In the ideal case of $\phi_s=\phi_i=\phi$, we define visibility for the $\phi$ basis as,
\begin{equation}
\label{eq:vis_general}
\mathcal{V}^{\phi}=|\frac{\mathcal{C}_{0,0}^\phi-\mathcal{C}_{0,1}^\phi-\mathcal{C}_{1,0}^\phi+\mathcal{C}_{1,1}^\phi}{\mathcal{C}_{0,0}^\phi+\mathcal{C}_{0,1}^\phi+\mathcal{C}_{1,0}^\phi+\mathcal{C}_{1,1}^\phi}|  
\end{equation} 

$\mathcal{C}_{x,y}^{\phi_s,\phi_i}$ is composed of true coincidences ($\mathcal{C}_{x,y}^{\phi_s,\phi_i,true}$), cross accidental coincidences between two consecutive photon-pairs ($\mathcal{C}_{x,y}^{\phi_s,\phi_i,cross-acc}$) and accidental coincidences due to the noisy background photons ($\mathcal{C}_{x,y}^{\phi_s,\phi_i, bkg-acc}$).

\begin{equation}
\label{coinc_total}
\mathcal{C}_{x,y}^{\phi_s,\phi_i}=\mathcal{C}_{x,y}^{\phi_s,\phi_i,true}+\mathcal{C}_{x,y}^{\phi_s,\phi_i,cross-acc}+\mathcal{C}_{x,y}^{\phi_s,\phi_i, bkg-acc}    
\end{equation}
True coincidences occur between the signal and the idler photon of the same photon pairs created from the source. For a detected coincidence event between the signal and the idler arm, the photon pair generated from the source must pass through the two polarizers simultaneously (see \figref{fig:Sagnac_theory_model}). Then the photon pair simultaneously must be detected in the single-photon detectors, amidst all the channel losses and noises. Hence, we assign probabilities for these two processes as the following,
\begin{equation}  \mathcal{C}_{x,y}^{\phi_s,\phi_i,true}=\mathcal{N}\{P_{x,y}^{1,1}\}\{q_{1,1}^{1,1}\}
\label{Truecoin}
\end{equation}

$\mathcal{N}$ pairs/sec is the brightness of the photon-pair source. $P_{x,y}^{a,b}$ is the probability that simultaneously $a$ number of photons pass through the polarizer in the signal arm while $(\phi_s,x)$ polarization direction is selected, and $b$ number of photons pass through the polarizer in the idler arm while $(\phi_i,y)$ polarization direction is selected; given the single photon-pair input state $\left|1\right\rangle_s\left|1\right\rangle_i$.
\begin{equation}
\label{eq:pol_projection1}
\begin{aligned}  
P_{x,y}^{a,b}&=Tr\{(\left|x\right\rangle_s^{\otimes a}\left|x^{\bot}\right\rangle_s^{\otimes(1-a)}\left\langle x\right|_s^{\otimes a}\left\langle x^{\bot}\right|_s^{\otimes(1-a)})\\
&\otimes(\left|y\right\rangle_i^{\otimes b}\left|y^{\bot}\right\rangle_i^{\otimes(1-b)}\left\langle y\right|_i^{\otimes b}\left\langle y^{\bot}\right|_i^{\otimes(1-b)})\rho_{s,i}\}
\end{aligned}
\end{equation}

for all $a\in\{0,1\}$ and $b\in\{0,1\}$. Note that $P_{x,y}^{1,1}=P_{x,y^{\bot}}^{1,0}=P_{x^{\bot},y}^{0,1}=P_{x^{\bot},y^{\bot}}^{0,0}$, $\forall~\phi_{s(i)}$. $q_{1,1}^{1,1}$ is the probability of detecting $1$ photon in both the signal and idler arm, given the $1$ output photon from both the polarizer setting. $q_{1,1}^{1,1}=\eta_s^{\prime}(\phi_s,x)\eta_i^{\prime}(\phi_i,y)$. Here, $\eta_{s(i)}^{\prime}(\phi_{s(i)},x(y))$ is the overall channel efficiency of the signal (idler) arm which might be sensitive to the polarization of the single photons. However, for the sake of simplicity, from now on we will only write $\eta_{s(i)}^{\prime}$.
\begin{equation}  \eta_{s(i)}^{\prime}=\eta_{s(i)}\eta_{s(i)}^\delta=\frac{\eta_{s(i)}}{1+\mathcal{N}_{s(i)}\delta_{s(i)}}
   \label{eq:etaprim}
\end{equation}

$\eta_{s(i)}$ is the polarization-sensitive overall channel efficiency when the detector deadtime $\delta_{s(i)}$ is zero. $\eta_{s(i)}^\delta$ reflects the additional loss due to the dead time of the detectors \cite{becares2012detector}. $\mathcal{N}_{s(i)}$ is the total detected single-photon count rate (including both signal and noise) in the individual arm for the $\delta_{s(i)}=0$ condition. $\mathcal{N}_s=\mathcal{N}\{P_{x,y}^{1,1}+P_{x,y}^{1,0}\}\eta_s+{\mathcal{I}}_{x}^{\phi_s}$. $\mathcal{N}_i=\mathcal{N}\{P_{x,y}^{1,1}+P_{x,y}^{0,1}\}\eta_i+{\mathcal{I}}_{y}^{\phi_i}$. Here ${\mathcal{I}}_{x(y)}^{\phi_{s(i)}}$ is the detected background noise in the signal (idler) arm when $\delta_{s(i)}=0$, for a fixed choice of the polarizer settings.

Cross-accidental coincidences usually occur in CW-pumped sources. Photon pairs are generated randomly in time inside the CW-pumped SPDC crystal. Due to the higher coincidence timing window compared to the single photon coherence time, it is possible to accidentally get a coincidence between a signal photon and an idler photon from different pairs.

\begin{equation}
\label{eq:cross-acc}
\begin{aligned}
\mathcal{C}_{x,y}^{\phi_s,\phi_i,cross-acc}=&\tau\{\mathcal{N}\sum_{a=0}^1\sum_{b=0}^1(P_{x,y}^{a,b})(q_{a,b}^{1,0})\}\\
&\times\{\mathcal{N}\sum_{a^{\prime}=0}^1\sum_{b^{\prime}=0}^1(P_{x,y}^{a^{\prime},b^{\prime}})(q_{a^{\prime},b^{\prime}}^{0,1})\}   
\end{aligned}   
\end{equation}

Here, $\tau$ is the coincidence window. The first and second curly brackets include the number of uncorrelated signal and idler photons respectively. $q_{a,b}^{1,0}$ is the probability of detecting $1$ signal photon and $0$ idler photon, $q_{a,b}^{0,1}$ is the probability of detecting $0$ signal photon and $1$ idler photon; given $(a,b)$ number of input photons.
\begin{equation}
\label{eq:Qfunction1}
   q_{a,b}^{1,0}=[1-\{1-\eta_s^{\prime}\}^a]\{1-\eta_i^{\prime}\}^b; q_{a,b}^{0,1}=\{1-\eta_s^{\prime}\}^a[1-\{1-\eta_i^{\prime}\}^b]
\end{equation}

For the pulse-pumped SPDC process, cross-accidental coincidences do not occur, as long as the coincidence or the gating window is much smaller than the pump pulse separation. 

$\mathcal{C}_{x,y}^{\phi_s,\phi_i, bkg-acc}$ term occurs due to the accidental coincidence between either one noisy background photon and one uncorrelated single photon from the source, or between two noisy photons. 
\begin{equation}
\label{eq:bkg-acc}
\begin{aligned}
\mathcal{C}_{x,y}^{\phi_s,\phi_i, bkg-acc}&=\mathcal{N}\sum_{a=0}^1\sum_{b=0}^1(P_{x,y}^{a,b})(q_{a,b}^{1,0}){\mathcal{I}}_{y}^{\phi_i}\eta_i^\delta\tau\\
&+\mathcal{N}\sum_{a^{\prime}=0}^1\sum_{b^{\prime}=0}^1(P_{x,y}^{a^{\prime},b^{\prime}})(q_{a^{\prime},b^{\prime}}^{0,1}){\mathcal{I}}_{x}^{\phi_s}\eta_s^\delta\tau\\
&+{\mathcal{I}}_{x}^{\phi_s}\eta_s^\delta{\mathcal{I}}_{y}^{\phi_i}\eta_i^\delta\tau
\end{aligned}
\end{equation}
In \eqref{eq:bkg-acc}, the first term is the accidental coincidence between a noisy signal and an uncorrelated idler photon, the second term is for the coincidence between a noisy idler and an uncorrelated signal photon, the third coincidence term is between one noisy signal and one noisy idler photon.

We consider some idealistic experimental approximations to simplify the equations for the coincidence count rate:
1) We optimize the polarization measurement basis in the signal and idler receiver module such that $\phi_s=\phi_i=\phi$. 
2) The overall channel efficiency (including detectors) is polarization insensitive, denoted as $\eta_{s(i)}^{\prime}$ for the signal (idler). 
3) Noisy background photons are overall unpolarized, so we use the fixed values $\mathcal{I}_x^{\phi_s}={\mathcal{I}}_s$ and $\mathcal{I}_y^{\phi_i}={\mathcal{I}}_i$. 
4) When dealing with the signal photon measurement in the noisy uplink channel and local measurement for the idler photon, noisy background photons are significantly higher in the signal channel compared to the idler channel. Therefore, we approximate ${\mathcal{I}}_i\approx0$, as we only consider the accidental coincidences due to the noisy photons in the signal arm. 
With these assumptions \eqref{coinc_total} becomes
\begin{equation}
\label{eq:coinc_sp_approx}
\begin{aligned}
\mathcal{C}_{x,y}^\phi&=\mathcal{N}P_{x,y}^{1,1}\eta_s^{\prime}\eta_i^{\prime}\\
&+\tau\mathcal{N}^2\eta_s^{\prime}\eta_i^{\prime}\{P_{x,y}^{1,0}+P_{x,y}^{1,1}(1-\eta_i^{\prime})\}\{P_{x,y}^{0,1}+P_{x,y}^{1,1}(1-\eta_s^{\prime})\}\\
&+\tau\mathcal{N}{\mathcal{I}}_s\eta_s^\delta\eta_i^{\prime}\{P_{x,y}^{0,1}+P_{x,y}^{1,1}(1-\eta_s^{\prime})\}
\end{aligned}   
\end{equation}

\textbf{Case 1: CW-pumped source}

\eqref{eq:coinc_sp_approx} applies to any CW-pumped entangled photon-pair source, given the assumptions mentioned earlier. We assume the spectral brightness of the source is $\mathcal{B}$ pairs/sec/mW/kHz. So, the number of entangled photon pairs created per second for $P_{pump}$ [mW] of pump power and $\Delta\omega$ [kHz] bandwidth is $\mathcal{N}=P_{pump}\mathcal{B}\Delta\omega$. We measure $\mathcal{N}$ or $\mathcal{B}$ directly by characterizing the source, using the simple equation $\mathcal{N}=(S_s\times S_i)/\mathcal{C}$ where $S_{s(i)}$ is the signal (idler) count rate and $\mathcal{C}$ is the coincidence count rate corrected for the accidental coincidences. Using \eqref{eq:coinc_sp_approx} in the visibility expression $\mathcal{V}^\phi$,
\begin{equation}
\label{eq:vis1X}   \mathcal{V}^{\phi}=\frac{\mathcal{V}_0^{\phi}\{1-\mathcal{N}\tau X_1-{\mathcal{I}}_s^{\prime}\tau\}}{1+\mathcal{N}\tau X_2+{\mathcal{I}}_s^{\prime}\tau X_3}
\end{equation}
Here we simplify by using polarization correlation as $P_{x,y}^{1,1}=P_c^\phi$ if $x=y$ and $P_{x,y}^{1,1}=P_{uc}^\phi$ if $x\neq y$, $\forall~\phi$. $\mathcal{V}_0^{\phi}$ is the polarization visibility of the photon-pair source itself in $\phi$ basis, defined as $\mathcal{V}_0^{\phi}=2(P_c^\phi-P_{uc}^\phi)$. $X_1=(\eta_s^{\prime}+\eta_i^{\prime}-\eta_s^{\prime}\eta_i^{\prime})/2$, $X_2=1-\eta_s^{\prime}/2-\eta_i^{\prime}/2+2\eta_s^{\prime}\eta_i^{\prime}\{(P_c^\phi)^2+(P_{uc}^\phi)^2\}$, $X_3=(2-\eta_s^{\prime})/\eta_s^{\prime}$. ${\mathcal{I}}_s^{\prime}$ is the detected noise rate in the signal arm for any projection of the polarizer; considering also the losses due to the dead time of the detector. 

In the experiment, we usually measure the average polarization visibility (APV) which is the average of at least two conjugate basis measurements.
\begin{equation}
\label{eq:APV}
    \text{APV}=\mathcal{V}^{avg}=(\mathcal{V}^{\phi=0}+\mathcal{V}^{\phi=\pi/4})/2
\end{equation}
Here $\phi=0$ represents the rectilinear basis, for which $x(y)=0$ and $x(y)=1$ is the horizontal ($0^{\circ}$) and vertical ($+90^{\circ}$) polarization axis, and $\phi=\pi/4$ represents the diagonal basis, for which $x(y)=0$ and $x(y)=1$ is the diagonal ($+45^{\circ}$) and anti-diagonal ($-45^{\circ}$) polarization axis.

\textbf{Case 2: pulse-pumped source} 

For the pulse-pumped entangled photon-pair sources, 
the second term in \eqref{eq:coinc_sp_approx} which is due to the cross-accidental coincidences, is negligibly small. Hence, the equation for visibility becomes $\mathcal{V}^{\phi}=\mathcal{V}_0^{\phi}(1-I_s^{\prime}\tau)/(1+I_s^{\prime}\tau X_3)$. We notice that the visibility for pulsed pumping is independent of the brightness of the source ($\mathcal{N}$), or the pumping power. However, this situation changes once we consider also the multi-photon pair generation from the source, which is non-negligible for higher pumping power.   

\subsection{SNR for the uplink communication channel}

In the literature, the polarization-visibility for an entangled photon-pair source is often represented as $\mathcal{V}=(signal-noise)/(signal+noise)$. This sets the minimum threshold for the SNR at $6:1$ for a secure QKD demonstration \cite{aspelmeyer2003long}, as the minimum requirement on the visibility is $1/\sqrt{2}$ for the violation of Bell's inequality. In terms of coincidence count rate, as defined in Sec. \ref{sec:General theoretical model}, $signal=\mathcal{C}_{0,0}^{\phi}+\mathcal{C}_{1,1}^{\phi}$ and $noise=\mathcal{C}_{0,1}^{\phi}+\mathcal{C}_{1,0}^{\phi}$. However, a drawback of this definition is its inability to distinguish between contributions from a non-ideal entangled-photon source and a lossy-noisy communication channel. For example, according to \eqref{eq:coinc_sp_approx}, both the $signal$ term and the $noise$ term have contributions from noisy background photons (${\mathcal{I}}_s$) in the uplink channel. Another approach to defining them is to consider $signal$ as only true coincidences, while $noise$ is all the accidental coincidences. This definition, although suitable for the pulse-pumped sources, fails for the CW-pumped condition, as the cross-accidental coincidences come from the source itself and not from noisy background photons. 

It is therefore intuitive to define SNR of the communication channels in the form of the function $f(\eta_{s(i)}^\prime/{\mathcal{I}}_{s(i)}^\prime)$ for the signal(idler) arm, where $\eta^\prime$ and ${\mathcal{I}}^\prime$ represents respectively the total efficiency of the communication channel and the detected noisy photon count rate coming from elsewhere. This representation is similar to defining SNR for the uplink/downlink channel in the case of non-entanglement-based QKD \cite{bonato2009feasibility}. For pulsed pumping, it is the ratio of the average number of signal photons to noisy photons per detected pulse \cite{er2005background}.

Here we define detected SNR for the uplink channel as $\text{SNR}_d=\eta_{s}^\prime/2{\mathcal{I}}_{s}^\prime\tau$. The numerator is the average probability for each signal photon created from the source (irrespective of their polarization) to reach the receiver module and getting detected; in the presence of noisy photons. The denominator is the probability of detecting a noisy background photon, in a polarization-independent way and the presence of signal photons. Note that we include a factor $2$ in the denominator as ${\mathcal{I}}_{s}^\prime$ is the average noise term for any polarization direction. $\text{SNR}_d$ can be measured directly from pre-characterizations, for both CW and pump-pulsed cases (see Sec. \ref{sec:Experimental Results}). $\mathcal{V}^{avg}$ in \eqref{eq:APV} can be written in terms of $\text{SNR}_d$ as the following.
\begin{equation}
\label{eq:vis3}   
\mathcal{V}^{avg}\approx\mathcal{V}_{0}^{avg}\times\frac{2-\mathcal{N}\tau\eta_i^\prime-\frac{\eta_s^\prime}{{\text{SNR}}_{d}}}{2+\mathcal{N}\tau(2-\eta_i^\prime)+\frac{2}{{\text{SNR}}_{d}}}
\end{equation}
\eqref{eq:vis3} applies to any CW pumped source. When it comes to pulsed pumping, the same equation can be used by disregarding the second terms in both the numerator and the denominator that include the brightness ($\mathcal{N}$). To derive \eqref{eq:vis3}, we further simplify \eqref{eq:vis1X} by using the approximation $P_c^{\phi}\approx 1/2$, which leads to $X_2\approx 1-X_1$. Additionally, we assume that $\eta_s^\prime<<\eta_i^\prime$, as the signal arm attenuation is much higher $(>30~\text{dB})$ after passing through a turbulent atmosphere. Therefore, we can approximately consider $X_1\approx \eta_i^\prime/2$ and $2-\eta_s^\prime\approx 2$. Here, $\mathcal{V}_{0}^{avg}$ represents the APV of the entangled photon source itself.

\subsection{Effect of the detection timing window, $\tau$ }
Until now, we have been assuming that the detection timing window is ideal for capturing all the signal and idler photons and only restricts the detection of noisy background photons. However, when the timing window ($\tau$) is much smaller than the temporal uncertainty of the single photons, the coincidence count rates include an additional efficiency factor, $\eta_\tau$, due to this timing window.

For this calculation, we assume that the temporal uncertainty distribution for both the signal (idler) photon is approximately Gaussian with the standard deviation $\sigma_{s(i)}$. Both of these values depend mostly on the timing jitter of the detectors and the accuracy of time synchronization. In the case of a pulse-pumped source, the signal (idler) detector is opened only for $\tau_{s(i)}$ time around the mean of the expected arrival time, triggered by either sharp reference pulses or by highly synchronized global clock systems. \eqref{eq:coinc_sp_approx} is modified to
\begin{equation*}
\label{eq:coinc_sp_approx_window_pulse}
\mathcal{C}_{x,y}^\phi=\mathcal{N}P_{x,y}^{1,1}\eta_s^{\prime\prime}\eta_i^{\prime\prime}+\tau\mathcal{N}{\mathcal{I}}_s\eta_s^\delta\eta_i^{\prime\prime}\{P_{x,y}^{0,1}+P_{x,y}^{1,1}(1-\eta_s^{\prime\prime})\}.
\end{equation*}
The only change here is that $\eta_{s(i)}^{\prime}$ is replaced by $\eta_{s(i)}^{\prime\prime}=\eta_{s(i)}^{\prime}\eta_{\tau}(\sigma_{s(i)})$, where 
\begin{equation*}
  \eta_{\tau}(\sigma_{s(i)})=\frac{1}{\sigma_{s(i)}\sqrt{2\pi}}\int_{-\tau/2}^{\tau/2}e^{\frac{-t^2}{2\sigma_{s(i)}^2}}dt.  
\end{equation*}
Visibility value is $\mathcal{V}^{\phi}=\mathcal{V}_0^{\phi}(1-I_s^{\prime}\tau)/(1+I_s^{\prime}\tau X_3^{\prime})$, where $X_3^{\prime}=(2-\eta_s^{\prime\prime})/\eta_s^{\prime\prime}$.

For the CW-pumped sources, no such timing reference is available. One can either apply the coincidence window in the post-processing stage or introduce a suitable time delay to the signal photon to allow the detector in the signal channel to be triggered based on local detection in the idler arm. In both cases, the temporal uncertainty of the coincidence profile is the convolution of individual temporal uncertainty of single photons, $\sigma_{T}\approx\sqrt{\sigma_s^2+\sigma_i^2}$. For the first scenario of post-processing, only the true coincidence term or the first term in \eqref{eq:coinc_sp_approx} is modified to $\mathcal{N}P_{x,y}^{1,1}\eta_s^{\prime}\eta_i^{\prime}\eta_{\tau}(\sigma_{T})$; leading to the modification in \eqref{eq:vis3}.
\begin{equation}
\label{eq:vis3_window}   
\mathcal{V}^{avg}\approx\mathcal{V}_{0}^{avg}\times\frac{2\eta_{\tau}(\sigma_{T})-\mathcal{N}\tau\eta_i^\prime-\frac{\eta_s^\prime}{{\text{SNR}}_{d}}}{2\eta_{\tau}(\sigma_{T})+\mathcal{N}\tau(2-\eta_i^\prime)+\frac{2}{{\text{SNR}}_{d}}}
\end{equation}
For the second scenario of idler-induced triggering, the modification in the coincidence count rate is slightly different, for example,
\begin{equation*}
\label{eq:coinc_sp_approx_window}
\begin{aligned}
\mathcal{C}_{x,y}^\phi&=\mathcal{N}P_{x,y}^{1,1}\eta_s^{\prime\prime}\eta_i^{\prime}\\
&+\tau\mathcal{N}^2\eta_s^{\prime}\eta_i^{\prime}\{P_{x,y}^{1,0}+P_{x,y}^{1,1}(1-\eta_i^{\prime})\}\{P_{x,y}^{0,1}+P_{x,y}^{1,1}(1-\eta_s^{\prime\prime})\}\\
&+\tau\mathcal{N}{\mathcal{I}}_s\eta_s^\delta\eta_i^{\prime}\{P_{x,y}^{0,1}+P_{x,y}^{1,1}(1-\eta_s^{\prime\prime})\}.
\end{aligned}   
\end{equation*}
Here, $\eta_s^{\prime\prime}=\eta_s^{\prime}\eta_{\tau}(\sigma_{T})$.
\subsection{Expected SNR for the uplink channel in the daylight condition}
The total signal attenuation for the ground-to-satellite uplink channel is represented by the channel efficiency parameter $\eta_{s}^\prime$, where the attenuation in $dB$ scale is given by $-10{\text{log}}_{10}(\eta_{s}^\prime)$. This total signal attenuation is a result of several factors, including attenuation in the sender module or ground station $(\eta_g)$, attenuation in the atmospheric channel due to absorption and scattering loss $(\eta_{atm})$, attenuation due to beam spreading in the turbulent atmosphere $(\eta_{bs})$, and losses in the satellite receiver module $(\eta_r)$, which includes detection efficiency and dead time of the detectors. Therefore, we can express $\eta_{s}^\prime=\eta_g\eta_{atm}\eta_{bs}\eta_r$.

The beam spreading in the uplink channel beam is mainly influenced by three processes: 1) diffraction-limited beam broadening of the Gaussian beam, 2) beam broadening due to small-scale turbulent effects, and 3) beam wandering due to large-scale turbulent effects. However, the beam-wandering can be minimized to a negligibly small amount by applying a state-of-the-art laser pointing and tracking system \cite{toyoshima2005long}. Therefore, the accuracy of the pointing mechanism plays a crucial role in limiting beam wandering. 
The beam waist radius ($W$) at the receiver module placed at a distance $L$ and $0^{\circ}$ zenith angle
from the sender module is \cite{dios2004scintillation}
\begin{equation}
\label{eq:beam_diameter}
    W^2(L)=W_{0}^2(1+\frac{L^2}{Z_{0}^2})+\frac{35.28L^2\gamma^2}{k^2r_{0}^2}+2{\left\langle \beta_{p}\right\rangle}^2
\end{equation}
$W_{0}$ is the beam waist radius at the sender telescope. $Z_0$ is the Rayleigh range; $Z_0=\pi W_0^2/\lambda$. $\gamma=1-0.26{(\frac{r_0}{W_0})}^{1/3}$. $k$ is the wave number, $k=2\pi/\lambda$. $\beta_{p}$ is the random displacement of the centroid of the beam due to the pointing error. $r_0$ is the Fried parameter calculated using the structure constant $(c_{n})$ of the refractive index in the atmosphere \cite{dios2004scintillation,andrews1995optical}.
\begin{equation}
\label{eq:r0}
    r_0={[0.42k^2\int_{0}^{L} c_{n}^2(z){(\frac{L-z}{L})}^{5/3} \,dz]}^{-3/5}
\end{equation}
\begin{equation}
\label{eq:Cn_value}
  \begin{aligned}
    c_{n}^2(z) & = 0.00594{(v/27)}^2{(z\times 10^{-5})}^{10} e^{(-z/1000)}\\
      & +2.7\times 10^{-16}e^{(-z/1500)}+Ae^{(-z/100)}
  \end{aligned}
\end{equation}
Here $A=1.7\times 10^{-14}\text{m}^{-2/3}$, $v=21~\text{m/s}$. Using Eq. (\ref{eq:r0},\ref{eq:Cn_value}) the approximate value of the Fried parameter is $r_0\approx 8.7~\text{cm}$, for the wavelength of $\lambda=785~\text{nm}$. Using \eqref{eq:beam_diameter}, the total signal attenuation in the uplink channel is \cite{bonato2009feasibility},
\begin{equation}
\eta_{s}^\prime(W_0,R,L)=\eta_g\eta_{atm}\{1-e^{-2R^2/W^2(L)}\}\eta_r 
\label{signalorbitcheff}
\end{equation}
$\eta_{s}^\prime$ is a function of beam waist radius at the sender output ($W_0$), radius of the receiver satellite ($R$), and distance of the satellite ($L$), assuming all the other parameters are constant.

During daylight, most of the background noise in the satellite uplink comes from the sunlight reflected off the Earth's surface. The expression for this noise is ${\mathcal{I}}_{day}=a_ER^2f^2H_{sun}$ \cite{bonato2009feasibility}, where $a_E$ represents the albedo of Earth ($a_E\approx 0.3$), $R$ is the radius of the receiving satellite, $f$ is the field of view (FOV) of the receiving satellite, and $H_{sun}$ is the spectral irradiance of the Sun at the Earth's surface. The detected noise count rate in the satellite receiver module (${\mathcal{I}}_s^\prime$) is given by ${\mathcal{I}}_s^\prime={\mathcal{I}}_{day}\times\eta_r\Delta\lambda_s/2$, where $\Delta\lambda_s$ is the spectral bandwidth of the filter placed in the receiver module, optimized for the spectral width of the signal photon from the source, and $\eta_r$ is the efficiency of the receiver module. The factor of $1/2$ is due to the polarization basis measurement.
Using both $\eta_s^\prime$ and ${\mathcal{I}}_s^\prime$,
\begin{equation}
\label{eq:SNR_detected}
\text{SNR}_d=\frac{\eta_s^\prime}{2{\mathcal{I}}_s^\prime\tau}=\frac{\eta_g\eta_{atm}\{1-e^{-2R^2/W^2(L)}\}}{a_ER^2f^2H_{sun}\Delta\lambda_s\tau}
\end{equation}
In the night-time scenario, the primary source of noise is the reflected moonlight from the Earth's surface, which is nearly six orders of magnitude lower than daylight,
${\mathcal{I}}_{night}\approx{\mathcal{I}}_{day}\times 10^{-6}$ \cite{bonato2009feasibility}.
 
\subsection{QKD applications}
\label{sec:QKD}
Entanglement-based QKD protocols require a minimum of two detectors in both the signal and idler channels in the case of active random basis selection. If passive random basis selection is used, four detectors are required in each channel. Although the general form of the coincidence count rate $\mathcal{C}_{x,y}^{\phi_s,\phi_i}$ remains similar to Eq. (\ref{coinc_total}-\ref{eq:bkg-acc}), the only modification comes to the expression of the uncorrelated signal (idler) count rates in \eqref{eq:cross-acc} and \eqref{eq:bkg-acc}. The expressions $\sum_{a=0}^1\sum_{b=0}^1(P_{x,y}^{a,b})(q_{a,b}^{1,0})$ and $\sum_{a^{\prime}=0}^1\sum_{b^{\prime}=0}^1(P_{x,y}^{a^{\prime},b^{\prime}})(q_{a^{\prime},b^{\prime}}^{0,1})$ are replaced by $(\sum_{m=\{y,y^\bot\}}P_{x,m}^{1,1})q_{1,1}^{1,0}$ and $(\sum_{m^\prime=\{x,x^\bot\}}P_{m^\prime,y}^{1,1})q_{1,1}^{0,1}$, respectively. Additionally, one must consider the contribution of more than two detectors clicking simultaneously. In such scenarios, it is necessary to select only one coincidence event uniformly randomly among all events \cite{lutkenhaus1999estimates, moroder2010entanglement}.

In a real QKD experiment, both Alice and Bob randomly choose between two measurement bases: rectilinear ($\phi_{s(i)}=0$) and diagonal ($\phi_{s(i)}=\pi/4$). This means that the probability of both choosing the same basis, say rectilinear, is $\frac{1}{4}$. This probability factor should be taken into account in the expression of the coincidence count rates as $\mathcal{C}_{x,y}^{\phi_s,\phi_i}\Rightarrow\mathcal{C}_{x,y}^{\phi_s,\phi_i}/4$.

For the selection of an active polarization basis, we can accurately model the experimental setup as shown in \figref{fig:Sagnac_theory_model}, where the polarizer setup directs the photon in one of the two perpendicular polarization directions for any chosen basis $\phi_{s(i)}$. Along with the two-fold coincidences, there is also a probability of three-fold and four-fold coincidences. Due to the uniform-random selection, we attribute a factor of $1/2$ and $1/4$ to the three-fold and four-fold cases. 
The modified coincidence count rate is
\begin{equation}
\label{eqn:qkdmodel1}
\begin{aligned}
&\mathcal{C}_{x,y}^{\phi_s,\phi_i}\Rightarrow\mathcal{C}_{x,y}^{\phi_s,\phi_i}\\
&+\frac{\tau}{2}\{\mathcal{C}_{x,y^\bot}^{\phi_s,\phi_i,true}\Gamma_{y}+\mathcal{C}_{x^\bot,y}^{\phi_s,\phi_i,true}\Gamma_{x}-\mathcal{C}_{x,y}^{\phi_s,\phi_i,true}(\Gamma_{x^\bot}+\Gamma_{y^\bot})\}\\
&+\frac{\tau}{4}\{\mathcal{C}_{x,y^\bot}^{\phi_s,\phi_i,true}\mathcal{C}_{x^\bot,y}^{\phi_s,\phi_i,true}-3\mathcal{C}_{x,y}^{\phi_s,\phi_i,true}\mathcal{C}_{x^\bot,y^\bot}^{\phi_s,\phi_i,true}\}
\end{aligned}
\end{equation}

In \eqref{eqn:qkdmodel1}, $\Gamma_{a}$ represents the combined count rate of the uncorrelated signal and noisy background photons in the channel labeled as $'a'$. The symbol $a$ belongs to the set $\{x, x^\bot\}$ for the signal channel and $\{y, y^\bot\}$ for the idler channel.
\begin{equation*}
\Gamma_{a}=\{\mathcal{N}(\sum_{m=\{y,y^\bot\}}P_{a,m}^{1,1})q_{1,1}^{1,0}\}+{\mathcal{I}}_{s}^{\prime},~~\forall a\in\{x,x^\bot\}
\end{equation*}
\begin{equation*}
\Gamma_{a^\prime}=\{\mathcal{N}(\sum_{m^\prime=\{x,x^\bot\}}P_{m^\prime,a^\prime}^{1,1})q_{1,1}^{0,1}\}+{\mathcal{I}}_{i}^{\prime},~~\forall a^\prime\in\{y,y^\bot\}
\end{equation*}

In the context of a passive random choice of the measurement basis, photons initially pass through a symmetric beamsplitter in both channels. This results in two output ports representing two conjugate bases: $\phi_s, \phi_s^{\dagger}$ in the signal channel and $\phi_i, \phi_i^{\dagger}$ in the idler channel. Consequently, in each channel, there are two polarizer setups and four detectors. As a result, this setup allows for a maximum of sixteen-fold coincidence events. A general modification to the coincidence count rate is detailed in \eqref{eqn:qkdmodel2}, where we only consider accidental coincidence terms up to one order of $\tau$, while neglecting higher orders with negligibly small probabilities of occurring.

\begin{equation}
\label{eqn:qkdmodel2}
\begin{aligned}
&\mathcal{C}_{x,y}\Rightarrow\mathcal{C}_{x,y}+\frac{\tau}{2}\{(\sum_{m\in\mathcal{Y}} {\mathcal{C}}^{true}_{x,m})\frac{\Gamma_{y}}{2}+(\sum_{m^\prime\in\mathcal{X}} {\mathcal{C}}^{true}_{m^\prime,y})\frac{\Gamma_{x}}{2}\\
&-\mathcal{C}_{x,y}^{true}(\sum_{n\in\mathcal{X}}\frac{\Gamma_{n}}{2}+\sum_{n^\prime\in\mathcal{Y}} \frac{\Gamma_{n^\prime}}{2})\}\\
&+\frac{\tau}{4}\{\sum_{g\in\mathcal{Y}} \sum_{g^\prime\in\mathcal{X}} \mathcal{C}_{x,g}^{true}\mathcal{C}_{g^\prime,y}^{true}-3\times\mathcal{C}_{x,y}^{true}(\sum_{h\in\mathcal{X}} \sum_{h^\prime\in\mathcal{Y}} \mathcal{C}_{h,h^\prime}^{true})\}
\end{aligned}
\end{equation}   

Here, $\mathcal{X}=\{x^\bot,x^\dagger,{(x^\dagger)}^\bot\}$ and $\mathcal{Y}=\{y^\bot,y^\dagger,{(y^\dagger)}^\bot\}$. $x^\dagger(y^\dagger)$ represents the conjugate polarization basis to $x(y)$. 
Another modification for the case of passive basis selection is in the expression of $\eta_{s(i)}^\delta$ (\eqref{eq:etaprim}). Here, one needs to replace $\mathcal{N}$ by $\mathcal{N}/2$ and ${\mathcal{I}}_{x(y)}^{\phi_{s(i)}}$ by ${\mathcal{I}}_{x(y)}^{\phi_{s(i)}}/2$, because of the symmetric beam splitter introduced in the experimental setup.

\subsection{Multi-photon scenarios}
 
When using relatively high pumping power, it is important to consider the probability of generating multiple photon pairs. These pairs with photon number $n>1$ can cause additional accidental coincidences, reducing the polarization visibility. In this section, we present a simple theoretical model to incorporate multi-photon pairs into our coincidence expression for the distribution of entanglement using only one detector in each receiver module. This model can easily be extended for QKD applications by following similar calculations outlined in section \ref{sec:QKD}.

We define the total coincidence count rate by summing up coincidences from all single-photon and multi-photon pairs with the joint number state $\left|n\right\rangle_s\left|n\right\rangle_i$. This can be expressed as $C_{x,y}^{\phi_s,\phi_i}=\sum_{n}C_{x,y,n}^{\phi_s,\phi_i}$, for all integer $n \geq 1$. We re-write here Eq. (\ref{coinc_total}-\ref{eq:bkg-acc}) with the inclusion of photon-number index $n$.
\begin{equation*}
C_{x,y,n}^{\phi_s,\phi_i}=C_{x,y,n}^{\phi_s,\phi_i,true}+C_{x,y,n}^{\phi_s,\phi_i,cross-acc}+C_{x,y,n}^{\phi_s,\phi_i, bkg-acc}    
\end{equation*}
\begin{equation*}  C_{x,y,n}^{\phi_s,\phi_i,true}=\mathcal{N}_n\sum_{a=0}^n\sum_{b=0}^n(P_{x,y,n}^{a,b})(q_{a,b}^{\bar{1},\bar{1}})  
\end{equation*}

\begin{equation*}
\label{eq:cross-acc_2}
\begin{aligned}
C_{x,y,n}^{\phi_s,\phi_i,cross-acc}=&\tau\{\mathcal{N}_n\sum_{a=0}^n\sum_{b=0}^n(P_{x,y,n}^{a,b})(q_{a,b}^{\bar{1},0})\}\\
&\times\{\mathcal{N}_n\sum_{a^\prime=0}^n\sum_{b^\prime=0}^n(P_{x,y,n}^{a^\prime,b^\prime})(q_{a^\prime,b^\prime}^{0,\bar{1}})\} 
\end{aligned}   
\end{equation*}

\begin{equation*}
\label{eq:bkg-acc_2}
\begin{aligned}
C_{x,y,n}^{\phi_s,\phi_i, bkg-acc}&=\tau\{\mathcal{N}_n\sum_{a=0}^n\sum_{b=0}^n(P_{x,y,n}^{a,b})(q_{a,b}^{\bar{1},0}){\mathcal{I}}_{y}^{\phi_i}\eta_i^\delta\\
&+\mathcal{N}_n\sum_{a^{\prime}=0}^n\sum_{b^{\prime}=0}^n(P_{x,y,n}^{a^{\prime},b^{\prime}})(q_{a^{\prime},b^{\prime}}^{0,\bar{1}}){\mathcal{I}}_{x}^{\phi_s}\eta_s^\delta\\
&+{\mathcal{I}}_{x}^{\phi_s}\eta_s^\delta{\mathcal{I}}_{y}^{\phi_i}\eta_i^\delta\}
\end{aligned}
\end{equation*}
$\mathcal{N}_n$ represents the number of $n$-photon pair events per unit time from the source, given by the equation $\mathcal{N}_n=\mathcal{N}_0p_n$. In this equation, $\mathcal{N}_0$ denotes the average pump photon rate interacting with the non-linear crystal inside the entire mode volume. The term $p_n$ denotes the photon number distribution from the source, similar to the distribution from a thermal source, and is given by the formula $p_n=\frac{\mu^n}{(1+\mu)^{n+1}}$, where $\mu$ is the average photon number, $\mu=\left\langle n\right\rangle$.
The probability $P_{x,y,n}^{a,b}$ represents the likelihood that $a$ number of photons pass through the polarizer in the signal arm with the '$x$' polarization direction selected, and $b$ number of photons pass through the polarizer in the idler arm with the '$y$' polarization direction chosen, given the input $\left|n\right\rangle_s\left|n\right\rangle_i$ photon number state on the polarizers. 
The terms $q_{a,b}^{\bar{1},\bar{1}}$, $q_{a,b}^{\bar{1},0}$, and $q_{a,b}^{0,\bar{1}}$ denote the probabilities of detecting at least $1$ photon in both the signal and idler arm, detecting at least $1$ signal photon and $0$ idler photon, and detecting $0$ signal photon and at least $1$ idler photon, respectively, given $(a,b)$ number of input photons just after the polarizers.
\begin{equation*}
   q_{a,b}^{\bar{1},\bar{1}}=[1-\{1-\eta_s^{\prime}(\phi_s,x)\}^a][1-\{1-\eta_i^{\prime}(\phi_i,y)\}^b] 
\end{equation*}
$q_{a,b}^{\bar{1},0}$ and $q_{a,b}^{0,\bar{1}}$ has the same expression as described in \eqref{eq:Qfunction1}.  $\tau$ is the coincidence window. $\eta_{s(i)}^{\prime}={\eta_{s(i)}}/(1+\mathcal{N}^{*}_{s(i)}\delta_{s(i)})$. Here $\mathcal{N}^{*}_{s(i)}$ is the total detected singles count rate contributed from all $n$-number of photon-pair states as well as background noise counts in the individual arm for $\delta_{s(i)}=0$ condition. $\mathcal{N}^{*}_{s}=\sum_n\sum_{a=1}^{n}\sum_{b=0}^{n}[\mathcal{N}_n\{1-(1-\eta_s^{\prime})^a\}P_{x,y}^{a,b}]+{\mathcal{I}}_{x}^{\phi_s}$, and $\mathcal{N}^{*}_{i}=\sum_n\sum_{a=0}^{n}\sum_{b=1}^{n}[\mathcal{N}_n\{1-(1-\eta_i^{\prime})^b\}P_{x,y}^{a,b}]+{\mathcal{I}}_{y}^{\phi_i}$.

The expression of $P_{x,y,n}^{a,b}$ for $n>1$ is more complex than what is shown in \eqref{eq:pol_projection1}. When dealing with multi-photon events, it may be beneficial to consider a Hamiltonian for an ideal maximally entangled source \cite{kok2000postselected, ma2007quantum, holloway2013optimal,neumann2021model}. Alternatively, we can use a different approach to reconstruct the polarization density matrices for higher-order multi-photon scenarios. In this alternative approach, we take into account more realistic source conditions rather than just idealistic ones.

We can generalize any SPDC-based polarization-entangled photon-pair sources as the following. Two near-identical sources ($c1$ and $c2$) creates the signal-idler polarization state $\left|H\right\rangle_s\left|H\right\rangle_i$ and $\left|V\right\rangle_s\left|V\right\rangle_i$, respectively with the collection ratio $|p_{c1}|^2:|p_{c2}|^2$. One may consider $c1,c2$ as two separate non-linear crystals placed side-by-side or parallel, or a single crystal doubly passed by the pumping light, or a single type-\Romannum{2} crystal creating two separate SPDC rings \cite{anwar2021entangled}. $|p_{c1}|^2:|p_{c2}|^2$ ratio may differ from ideal $1:1$ depending on the asymmetric pumping condition, geometry, alignment of the source setup, and asymmetric collection efficiency. Both of these states then travel through the same or different optical paths depending on the setup before they merge. Different optical paths can introduce different polarization rotations of the initial two-photon states. In the case of the non-degenerate SPDC process, separate wavelengths of the signal and idler photons may also affect the polarization rotation differently. The next step is to merge the two states in such a way that they are indistinguishable in terms of spectral, spatial, and temporal characteristics. However, for a realistic implementation, we assign an indistinguishability parameter $\alpha$, where $0\leq\alpha\leq1$.

Let us consider a scenario involving multi-photon pairs, where photon pairs are created in the Fock state $\left|n\right\rangle_s\left|n\right\rangle_i$. This state can be created from any of the $(n+1)$ possible pumping scenarios. For instance, when $n=2$, the state can be created in three different ways: source $c1$ creates both photon pairs $\left|2\right\rangle_s\left|2\right\rangle_i$ by annihilating $2$ pump photons simultaneously, source $c1$ and $c2$ both create $\left|1\right\rangle_s\left|1\right\rangle_i$ simultaneously, or source $c2$ creates both photon pairs $\left|2\right\rangle_s\left|2\right\rangle_i$. 
Considering the SPDC operations, $c1\xrightarrow{\text{SPDC}} \left|H\right\rangle_s\left|H\right\rangle_i$ and $c2\xrightarrow{\text{SPDC}} \left|V\right\rangle_s\left|V\right\rangle_i$, we write the three possible initial polarization states as $\left|H\right\rangle_s^{\otimes 2}\left|H\right\rangle_i^{\otimes 2}$, $\left|H\right\rangle_s\left|H\right\rangle_i\left|V\right\rangle_s\left|V\right\rangle_i$, and $\left|V\right\rangle_s^{\otimes 2}\left|V\right\rangle_i^{\otimes 2}$. Similarly we write all $(n+1)$ possible initial polarization states for any value of $n$ as $\left|\psi_{in,n}\right\rangle_k=\left|H\right\rangle_s^{\otimes k}\left|H\right\rangle_i^{\otimes k}\left|V\right\rangle_s^{\otimes (n-k)}\left|V\right\rangle_i^{\otimes (n-k)}$, for $k=0,1,2,..,n$. Each state $\left|\psi_{in,n}\right\rangle_k$ has probability $|\mathcal{P}_{n,k}|^2=|p_{c1}^kp_{c2}^{(n-k)}|^2/{\sum_k |p_{c1}^kp_{c2}^{(n-k)}|^2}$, where $p_{c1},p_{c2}$ may include also complex phase term coming from pumping condition, and $|p_{c1}|^2+|p_{c2}|^2=1$.

Polarization rotation operation that may depend on the path, as well as the signal (idler) frequencies, is
\begin{equation*}
   U_{PR}:\left|H\right\rangle_{s(i)}\Rightarrow cos\theta_{s(i)}\left|H\right\rangle_{s(i)}+sin\theta_{s(i)}e^{i\phi_{s(i)}}\left|V\right\rangle_{s(i)} 
\end{equation*}
\begin{equation*}
   :\left|V\right\rangle_{s(i)}\Rightarrow cos\theta_{s(i)}^{'}\left|V\right\rangle_{s(i)}+sin\theta_{s(i)}^{'}e^{i\phi_{s(i)}^{'}}\left|H\right\rangle_{s(i)} 
\end{equation*}

Finally, the density matrix represented for $n$ number of photon-pair is,
\begin{equation}
\label{eq:multi_density_matrix}
\rho_{n}=\alpha\left|\psi_{pure,n}\right\rangle\left\langle\psi_{pure,n}\right|+(1-\alpha)\rho_{\text{mix},n}
\end{equation}
$\left|\psi_{pure,n}\right\rangle$ state is created for perfectly indistinguishable photon-pairs from $c1$ and $c2$. 
$\left|\psi_{pure,n}\right\rangle=U_{PR}^{\otimes 2n} (\sum_{k=0}^{n}\mathcal{P}_{n,k}\left|\psi_{in,n}\right\rangle_{k})=\sum_{k1=0}^{n}\sum_{k2=0}^{n}\mathcal{P}^{*}_{n,k1,k2}\left|\psi_{in,n}^{*}\right\rangle_{k1,k2}$ with state re-normalization such that $\sum_{k1=0}^{n}\sum_{k2=0}^{n}|\mathcal{P}^{*}_{n,k1,k2}|^2=1$. Note that for the indistinguishable case, only the $(n+1)^2$ states $\left|\psi_{in,n}^{*}\right\rangle_{k1,k2}$ span the whole polarization basis set; $\left|\psi_{in,n}^{*}\right\rangle_{k1,k2}=\left|H\right\rangle_s^{\otimes k1}\left|H\right\rangle_i^{\otimes k2}\left|V\right\rangle_s^{\otimes (n-k1)}\left|V\right\rangle_i^{\otimes (n-k2)}$. For example, in case of $n=2$ condition, all the polarization states $\left|H\right\rangle_s\left|H\right\rangle_i\left|V\right\rangle_s\left|V\right\rangle_i$, $\left|V\right\rangle_s\left|H\right\rangle_i\left|H\right\rangle_s\left|V\right\rangle_i$, $\left|V\right\rangle_s\left|V\right\rangle_i\left|H\right\rangle_s\left|H\right\rangle_i$ and $\left|H\right\rangle_s\left|V\right\rangle_i\left|V\right\rangle_s\left|H\right\rangle_i$ are indistinguishable from each other; hence written in only one basis state $\left|H\right\rangle_s\left|H\right\rangle_i\left|V\right\rangle_s\left|V\right\rangle_i$.

On the other hand, for completely distinguishable cases, all possible $2^{2n}$ polarization states span the basis set for the mixture $\rho_{\text{mix},n}$. Here $\rho_{\text{mix},n}=\sum_{k=0}^{n}|\mathcal{P}_{n,k}|^2\rho_{\text{mix},n}^{k}$. We define $\rho_{\text{mix},n}^{k}=\left|\psi_{in,n}^{norm}\right\rangle_{k}\left\langle\psi_{in,n}^{norm}\right|_{k}$, where $\left|\psi_{in,n}^{norm}\right\rangle_{k}$ is a re-normalized state of $U_{PR}^{\otimes 2n} \left|\psi_{in,n}\right\rangle_{k}$. Hence, using \eqref{eq:multi_density_matrix} we write the general expression for $P_{x,y,n}^{a,b}$ as the following, 
\begin{widetext}
\begin{equation*}
\begin{aligned}
&P_{x,y,n}^{a,b}=\alpha~Tr\{(\left|x\right\rangle_s^{\otimes a}\left|x^{\bot}\right\rangle_s^{\otimes(n-a)}\left\langle x\right|_s^{\otimes a}\left\langle x^{\bot}\right|_s^{\otimes(n-a)})\otimes(\left|y\right\rangle_i^{\otimes b}\left|y^{\bot}\right\rangle_i^{\otimes(n-b)}\left\langle y\right|_i^{\otimes b}\left\langle y^{\bot}\right|_i^{\otimes(n-b)})\left|\psi_{pure,n}\right\rangle\left\langle\psi_{pure,n}\right|\}\\
&+(1-\alpha)\sum_{k=0}^{n}|\mathcal{P}_{n,k}|^2\sum_{c=c_{min}}^{c_{max}}\sum_{c^{\prime}=c^{\prime}_{min}}^{c^{\prime}_{max}}Tr\{(\left|x\right\rangle_{s1}^{\otimes c}\left|x^{\bot}\right\rangle_{s1}^{\otimes (k-c)}\left|x\right\rangle_{s2}^{\otimes (a-c)}\left|x^{\bot}\right\rangle_{s2}^{\otimes(n-a-k+c)}\left\langle x\right|_{s1}^{\otimes c}\left\langle x^{\bot}\right|_{s1}^{\otimes (k-c)}\left\langle x\right|_{s2}^{\otimes (a-c)}\\
&\left\langle x^{\bot}\right|_{s2}^{\otimes(n-a-k+c)})\otimes(\left|y\right\rangle_{s1}^{\otimes c^{\prime}}\left|y^{\bot}\right\rangle_{s1}^{\otimes (k-c^{\prime})}\left|y\right\rangle_{s2}^{\otimes (b-c^{\prime})}\left|y^{\bot}\right\rangle_{s2}^{\otimes(n-b-k+c^{\prime})}\left\langle y\right|_{s1}^{\otimes c^{\prime}}\left\langle y^{\bot}\right|_{s1}^{\otimes (k-c^{\prime})}\left\langle y\right|_{s2}^{\otimes (b-c^{\prime})}\left\langle y^{\bot}\right|_{s2}^{\otimes(n-b-k+c^{\prime})})\rho_{\text{mix},n}^{k}\}
\end{aligned}
\end{equation*}
\end{widetext}
We use $c_{min}=max\{0,a+k-n\}$, $c_{max}=min\{a,k\}$, $c^{\prime}_{min}=max\{0,b+k-n\}$, $c^{\prime}_{max}=min\{b,k\}$. To calculate all the probabilities $P_{x,y,n}^{a,b}$ we reconstruct all relevant density matrices $\left|\psi_{pure,n}\right\rangle\left\langle\psi_{pure,n}\right|$ and $\rho_{mix,n}^k$, numerically from the single experimentally reconstructed density matrix $\rho_{s, i}$. While measuring for $\rho_{s, i}$ the pump power is decreased to a low value ($\approx \mu\text{W}$) such that the multi-photon pair generation probability becomes negligibly small. For such an approximated condition, the density matrix for single photon ($n=1$) pair generation ($\rho_1$ from \eqref{eq:multi_density_matrix}) is numerically fitted with $\rho_{s,i}$. We apply the minimization of the trace distance between $\rho_1$ and $\rho_{s,i}$, $T(\rho_1,\rho_{s,i})=\frac{1}{2}Tr\{\sqrt{(\rho_1-\rho_{s,i})^{\dagger}(\rho_1-\rho_{s,i})}\}$ for the estimation of the optimal value of all the fitting parameters; $|p_{c1}|^2$, $arg(p_{c2})-arg(p_{c1})$, $\theta_{s(i)}$, $\theta_{s(i)}^{'}$, $\phi_{s(i)}$, $\phi_{s(i)}^{'}$, and $\alpha$. Using these parameters, we then reconstruct all $\rho_{n}$ matrices for $n>1$.
\begin{figure}[h!]
\centering
\includegraphics[height=5.5cm]{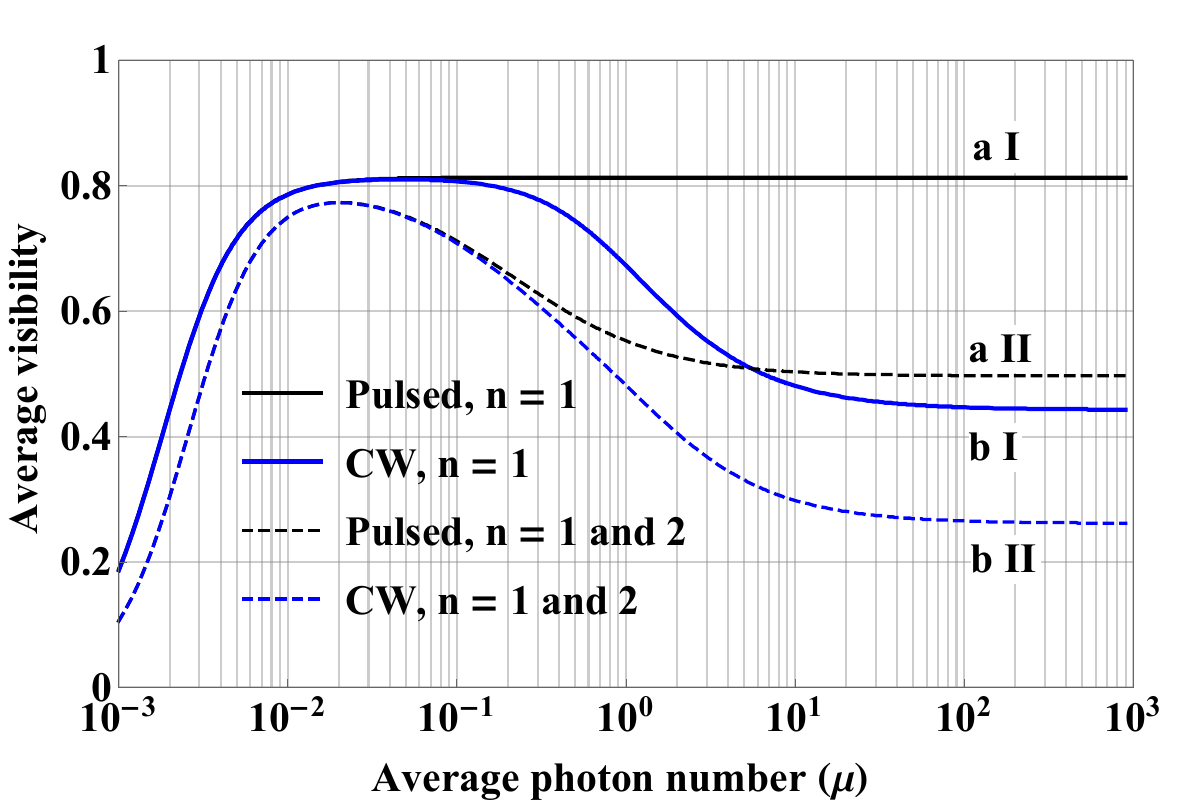}
\caption{Plot of the average visibility as a function of average photon number ($\mu$) for four scenarios: pulse-pumped (a) and CW-pumped (b) source considering only single-photon (\Romannum{1}) and up to two-photon (\Romannum{2}) pairs. }
\label{fig:SagSattMulti}
\end{figure}

In \figref{fig:SagSattMulti}, we illustrate the impact of multi-photon pair generation (up to $n=2$) on the APV under both CW and pulsed-pumping conditions. The x-axis represents the average photon number ($\mu$), which is proportional to the pumping power. In the CW case, pumping power is defined as the average pump power. However, in the pulsed case, it is the average pump power per pulse. It is important to note that in a pulse-pumped source, the total pumping power equals the average pump power per pulse multiplied by the pulse repetition rate. Therefore, one can safely increase the total pump power by simply increasing the pulse repetition rate. Increasing the pulse repetition rate does not affect the APV, as long as the separation between two consecutive pulses is larger than the coincidence window. 
When we disregard the occurrence of multi-photon pairs, the APV in the pulsed case remains unchanged as the pump power increases. However, in the CW case, the APV decreases due to accidental coincidences between consecutive single-photon pairs. If we also take into account the generation of two-photon pairs, the APV decreases with increasing pump power for both CW and pulsed cases. This is because higher pump power leads to a greater number of two-photon pair events, resulting in more accidental coincidences. Conversely, reducing the pump power (or $\mu$) also decreases APV for all four cases shown in the plot. This occurs because lowering the pump power reduces the signal photons or the brightness of the source, while the background noise level remains constant. As a result, the SNR decreases, leading to reduced APV. 

\section{Optical setup}
\label{sec3}
\begin{figure*}
\centering
\includegraphics[width=0.9\textwidth]{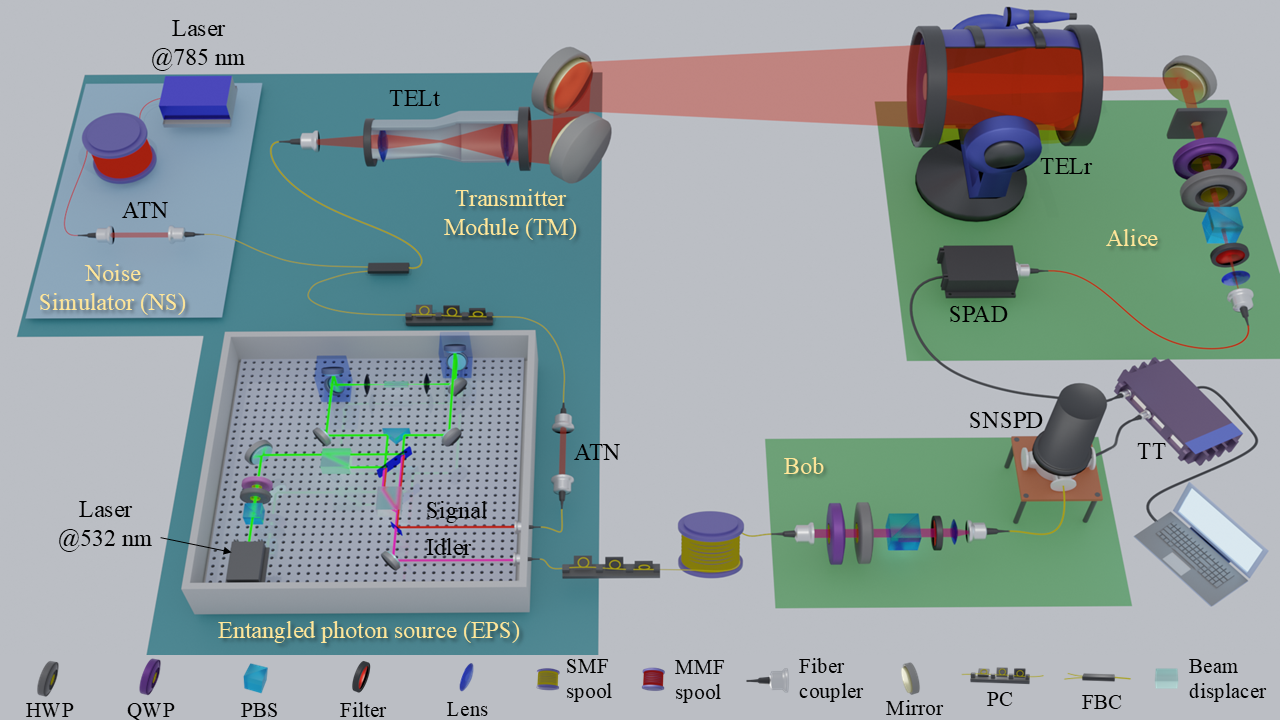}
\caption{Schematic of the experimental setup. ATN: tunable attenuation module consists of two fiber couplers; TELt: transmitter telescope; TELr: receiver telescope; SPAD: single photon avalanche detector; SNSPD: superconducting nanowire single photon detector; TT: time tagging device; HWP: half waveplate; QWP: quarter waveplate; PBS: polarizing beamsplitter; SMF: single mode fiber; MMF: multi-mode fiber; PC: polarization controller; FBC: fiber beam combiner.   }
\label{fig:satt3D}
\end{figure*}

Our experimental setup includes a source of polarization-entangled photon pairs (EPS) with high fidelity and pair generation rate. Each pair consists of a signal photon and an idler photon. The signal photon is directed to the transmitter module (TM), which sends it to the receiver module (Alice) through a free-space channel. Meanwhile, a fiber channel directs the idler photon to another receiver (Bob). Each receiver module is equipped with a single-photon detection system and an active polarization analyzer.
To simulate the daylight-uplink quantum communication channel, we have incorporated a noise simulator (NS) module into the TM to replicate the effects of background noise, particularly in daylight conditions. A tunable attenuation module (ATN) is also positioned between the EPS and the TM to simulate signal photon loss in the ground-to-satellite uplink channel. The schematic of our experimental setup is presented in \figref{fig:satt3D}.

\subsection{Entangled photon-pair source}
The polarization-entangled photon pairs are generated by the entangled photon-pair source (EPS) using a type-0 nonlinear crystal (MgO:PPLN) within a Saganac interferometer, similar to the setup described in Ref. \cite{szlachetka2023ultrabright}. The resulting state from the source is ideally represented as $(\left|H\right\rangle_{s}\left|H\right\rangle_{i}-\left|V\right\rangle_{s}\left|V\right\rangle_{i})/\sqrt{2}$.
These photon pairs are created via the SPDC process, where a CW pump laser light with a central wavelength of $532~\text{nm}$ is converted into signal photons at $785~\text{nm}$ and idler photons at $1651~\text{nm}$. The choice of $785~\text{nm}$ for the signal wavelength is ideal for uplink scenarios \cite{Bourgoin_2013}. Additionally, the idler wavelength falls within the U-band range of telecommunication, enabling long-range fiber connections \cite{valivarthi2016quantum}.
Moreover, the signal photon has a root-mean-squared spectral width of $0.4~\text{nm}$, facilitating precise spectral filtering in the receiver module for enhanced noise cancellation. The brightness of the source is $8.6(1)\times10^6~\text{pairs/sec/mW}$, with an APV of $0.97(1)$ and concurrence of $94.7(2)\%$.

Signal photons originating from the EPS are directed into an SMF. Before reaching the transmitter module (TM), they pass through an ATN and are joined with the noisy photons from the NS using a fiber beam combiner (FBC).
ATN consists of two fiber couplers, decoupling the photons into the free space and then recoupling them into an SMF. Precise attenuation of the signal photons is achieved by finely adjusting the second coupler. During the experiment, we regulate the overall attenuation of the signal channel between $20~\text{dB}$ and $50~\text{dB}$ solely through adjustments to the ATN. This attenuation level effectively replicates photon loss experienced in the uplink atmospheric channel between ground stations and LEO satellites.
Following the ATN, a polarization controller (PC) corrects the polarization of the signal photons according to the projective measurement basis on the receiver module (Alice).

\subsection{Background noise simulator}
The background noise simulator (NS) comprises a pulsed laser (Chameleon Ultra II) operating at a repetition rate of $80~\text{MHz}$. This laser is configured to emit light at a central wavelength of $785~\text{nm}$, with a linewidth (full width at half maximum) of $4.62~\text{nm}$. The choice of the central wavelength aligns with the signal photons emitted from the EPS.
We employ a two-step attenuation process to match the laser intensity with the background noise level in the uplink channel. Initially, a neutral density filter with fixed attenuation is utilized, followed by an additional adjustment using an ATN module to fine-tune the noisy photon count rate. This enables precise control over the detected background noise, ranging from around $100~\text{Hz}$ to a few MHz.
Stray photons detected by the satellite originate from various sources such as the sun, moon, city lights, blackbody radiation from the earth, and other stars. Hence, statistically these detected photons are unpolarized. To simulate this unpolarized nature, we pass the photons emitted by the NS through a $20~\text{m}$ long multi-mode fiber (MMF). However, this process only achieves a degree of polarization (DOP) of $50~\%$. Consequently, we manually adjust the ATN for each polarization analysis conducted in the Alice module to ensure uniform detection of noisy photon counts across all polarizations.

\subsection{Transmitter module}

Signal photons originating from the EPS and noisy photons from the NS undergo transmission through the transmitter module (TM). For this purpose, we employ an SMF-based FBC with a splitting ratio of $70:30$. The EPS photons are coupled to one input arm of the FBC, which has an input-to-output transmission efficiency of $70\%$. Signal and noise photons, sharing identical wavelengths and thus coupled to the same SMF, ensure indistinguishability in the spatial mode and equal collection efficiency in the receiver module.

The TM comprises a fiber package collimator, which decouples and collimates the incoming photons from the SMF to free space. Subsequently, the photons traverse through a telescope (TELt) comprised of two lenses with respective focal lengths of $50~\text{mm}$ and $400~\text{mm}$, resulting in an expanded output beam diameter (FWHM) of approximately $17~\text{mm}$. These photons are then directed to the receiver module (Alice) via two adjustable tip-tilt mirrors, offering precise control over the beam's direction.

The light transmitted from the TM travels a distance of $10~\text{m}$ to reach the receiver module, Alice. The free-space channel is established within a closely controlled laboratory environment, maintaining the air temperature at $21\pm0.5$ \textcelsius ~using a laminar flow system.

\subsection{Receiver module (Alice and Bob)}
The receiver module (Alice) comprises a Schmidt-Cassegrain $f/10$ telescope (TEL), an active polarization analyzer, and a SPAD detector for single photon detection. TEL is a commercial model (Meade) with an aperture size of $254~\text{mm}$ and a focal plane of $2.5~\text{m}$. The signal photons captured by the TEL first pass through an array of pinholes carefully aligned to pass only the central Gaussian spatial mode and block any ambient light from entering the optical path \cite{peloso2009daylight}. Next, the polarization analyzing measurement is performed by using a motorized (Standa 8MPR16-1) quarter-waveplate (QWP), a motorized half-waveplate (HWP), and a polarizing beamsplitter (PBS). Additionally, a narrow-bandpass filter, F1  (Semrock LL01-785-12.5) with a central wavelength of $785~\text{nm}$ and spectral width (RMS) of $1~\text{nm}$ is integrated into this receiver module. All the spectrally filtered photons are coupled to an MMF using a fiber coupler (FC), allowing for the coupling of approximately $50~\%$ of photons sent from the TM. These coupled photons are then detected by a silicon-based SPAD detector (Excelitas Technologies SPCM-AQRH-14-FC) with a detection efficiency of $0.5$, $32~\text{ns}$ dead time, and $100~\text{cps}$ dark count rate. 

In our experimental setup, all the optical components in the receiver module including the polarization analyzer, spectral filter, and fiber couplers are attached to the telescope using an opto-mechanical caging structure (not shown in \figref{fig:satt3D}). This setup aids in alignment and ensures stable coupling efficiency. The entire cage is shielded from scattered light by a custom-made black box. Furthermore, the entire receiver system, including the telescope, is covered with a black cloth to prevent light from entering from the sides.
To assess the effectiveness of our light-blocking strategy, we initially cover the telescope aperture and measure the background photon count rate caused by light entering from the sides in a fully illuminated laboratory setting. The noise count rate is detected in the order of $100~\text{cps}$, the same as the SPAD dark count rate.

The idler photons of wavelength $1651~\text{nm}$ from the EPS are sent to the receiver module Bob via an SMF. For the local measurement of the APV, we choose a shorter SMF length of $1~\text{m}$, while for the final QKD demonstration, we employ a $10~\text{km}$ long SMF fiber spool. Another PC is also used for the polarization correction in the idler arm. The Receiver module (Bob) comprises a fiber package collimator, polarization analyzer including QWP, HWP, and PBS, and a long-pass filter (F2, ThorLabs FELH-1500). Idler photons are then coupled into an SMF connected to a superconducting nanowire detector (SNSPD). This SNSPD from Scontel has a detection efficiency of $70 \%$, $1000~\text{cps}$ dark count rate, and a timing jitter (FWHM) of $34~\text{ps}$.

\subsection{Coincidence measurement and data analysis}
The SPAD detector in the Alice module generates a TTL signal upon photon detection. This voltage signal, with a pulse timing width of $10~\text{ns}$ and an amplitude of $2.2~\text{V}$, travels via an SMA cable to the time-to-digital converter (time tagger, TT). Simultaneously, the Bob SNSPD detector produces a short voltage pulse upon photon detection. This pulse, with an amplitude of $326~\text{mV}$ and a fall time (from $90~\text{\%}$ to $10~\text{\%}$) of 17 ns, is also sent via an SMA cable to the same TT device. At the TT, once the incoming voltage pulse reaches a set triggering level, it records the time of photon arrivals relative to its internal clock. The trigger at the TT is configured to respond to the rising edge of the voltage pulse, with a threshold set at 1 V for Alice's signal and 180  mV for Bob's signal. This allows us to perform time correlation analysis by producing histograms displaying the time differences between the arrivals of photons from Alice and Bob on the x-axis. The y-axis represents the occupancy of the time slots. The histogram bins are set to 1 ps, whereas the TT device introduces a timing jitter of $10~\text{ps}$. Due to the SPDC process in our EPS, we observe strong correlations in the time domain. This enables us to apply precise timing filtering. However, due to the relatively higher timing jitter of the used detector compared to the coherence time of the single photons, we choose the coincidence window as $1~\text{ns}$.

\section{Verification Results}
\label{sec:Experimental Results}
We begin by measuring the entanglement distribution between Alice and Bob. To do this, we measure the visibility of polarization in two separate conjugate bases (rectilinear and diagonal). For a secure QKD demonstration, the APV must be at least $71\%$, which is the minimum threshold for violating the Bell-CHSH inequality \cite{aspelmeyer2003long}.
 
\begin{figure}[h!]
\centering
\includegraphics[height=5.5cm]{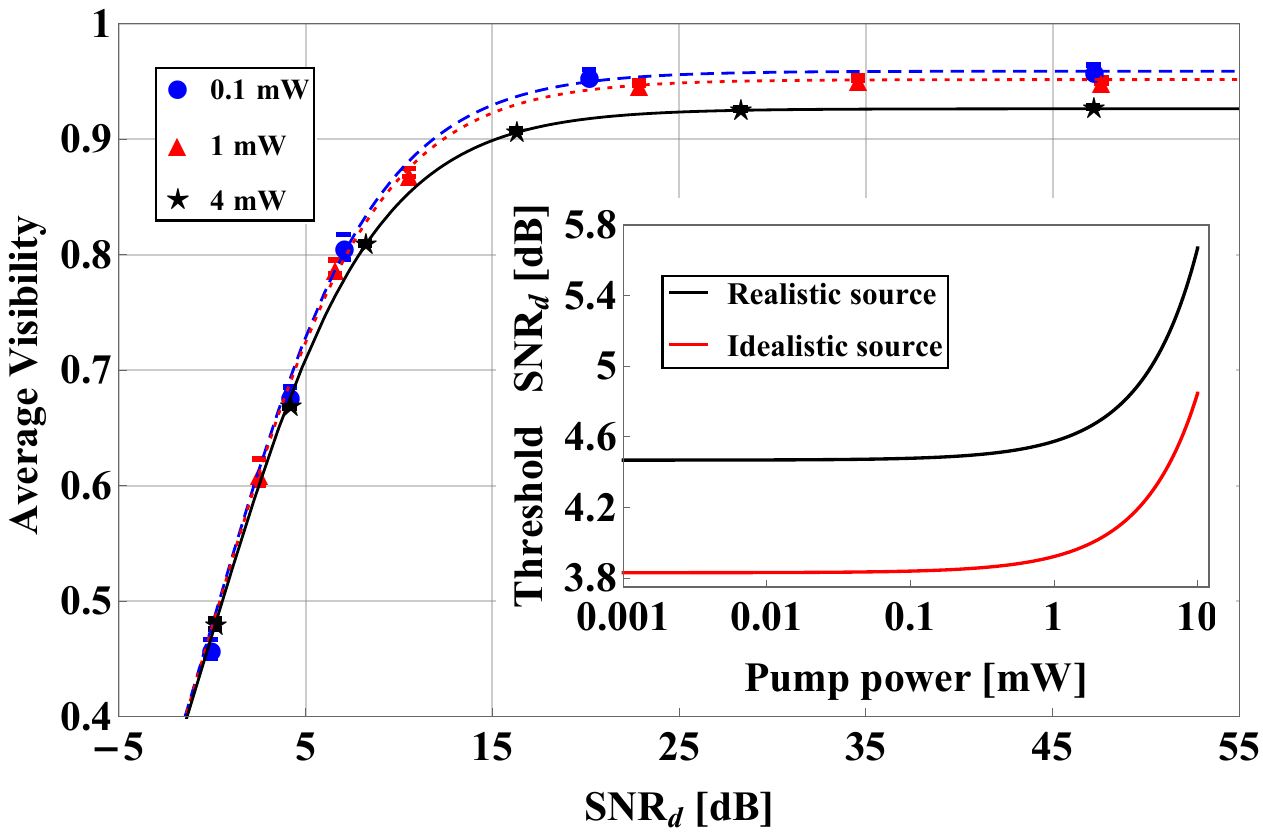}
\caption{Average visibility as a function of $\text{SNR}_d$, for the uplink channel. Experimental data for Three different powers of the CW pump laser is plotted along with theoretical predictions obtained from \eqref{eq:vis3}. The inset plot shows threshold $\text{SNR}_d$ versus pump power for both ideal and realistic sources.}
\label{fig:Vis_vs_SNRd}
\end{figure}

We collect three sets of visibility data for three different source brightness levels by adjusting our CW laser pumping power to $0.1~\text{mW}$, $1~\text{mW}$, and $4~\text{mW}$. 
For each brightness level, we measure the visibility (defined in \eqref{eq:vis_general}) in the rectilinear and diagonal basis for different values of ${\text{SNR}}_d$. We adjust the ${\text{SNR}}_d$ by adjusting both the signal attenuation ($\eta_s^\prime$) and the background noise level ($I_s^\prime$). We measure $\eta_s^\prime$ and $I_s^\prime$ directly from the experiment 
as $\eta_s^\prime=2S_{sig}/\mathcal{N}$, $I_s^\prime=S_{noise}$. Here, $S_{sig}$ represents the average detected signal photon count rate for any polarization projection on the receiver module, with no additional background noise added and the dark count rate of the detector subtracted. $\mathcal{N}$ represents the brightness of the source. $S_{noise}$ stands for the average detected noisy photon count rate, including the dark count rate, for any polarization projection on the receiver module. 

In \figref{fig:Vis_vs_SNRd}, we observe that the APV drops quickly for a low ${\text{SNR}}_d$ value, while it saturates as the ${\text{SNR}}_d$ value increases. Additionally, we observe a slight decrease in the overall APV profile as the pump power increases. These experimental observations align well with the predictions from our theoretical model. It is important to note that the theory plots shown in \figref{fig:Vis_vs_SNRd} are not numerical fits to the experimental data, but rather they represent theoretically expected data generated from \eqref{eq:vis3} using experimentally measured parameters from our setup: $\mathcal{V}_{0}^{avg}=0.96$, $\mathcal{N}=8.6\times10^6~\text{counts/sec/mW}$, $\eta_i^\prime=16~\text{dB}$ of attenuation, $\tau=1~\text{ns}$, $\eta_s^\prime=30~\text{dB}$ of attenuation. The inset graph illustrates the relationship between the CW pump power and the threshold ${\text{SNR}}_d$ value for violating the Bell-CHSH inequality. In this study, with the near-ideal polarization-entangled source, the threshold ${\text{SNR}}_d$ is approximately $4.5~\text{dB}$. With an ideal source, this threshold can be reduced to $3.8~\text{dB}$.

\begin{figure}[h!]
\centering
\includegraphics[height=5.5cm]{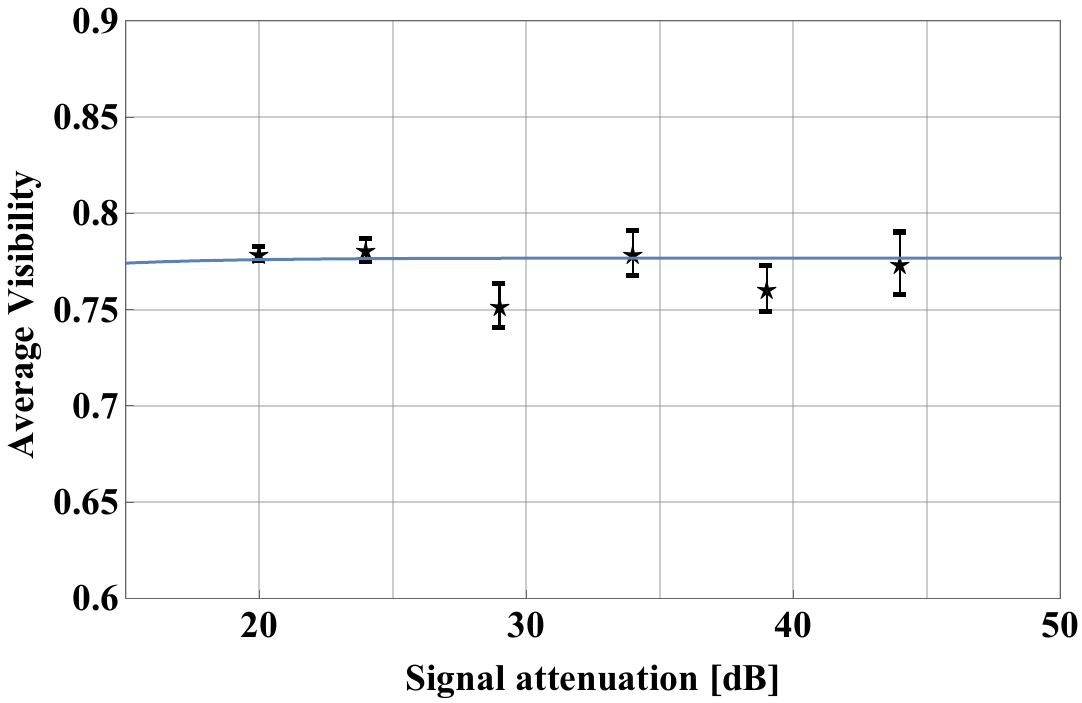}
\caption{Average visibility as a function of signal channel attenuation, $\eta_s^\prime$, while the detected noise photon count rate, ${\mathcal{I}}_s^\prime$, is adjusted for each point such that the $\text{SNR}_d$ value is always fixed at $4.6~\text{dB}$.}
\label{fig:Vis_vs_SA}
\end{figure}

One can deduce from \eqref{eq:vis3} that the APV depends on the $\text{SNR}_d$ and not on the individual $\eta_s^\prime$ and $I_s^\prime$ parameters. This is assuming that $\eta_s^\prime$ is significantly smaller than $\text{SNR}_d$, which is true for the uplink scenario being considered. To experimentally verify this, we measure the APV at different levels of signal attenuation ($\eta_s^\prime$), while keeping the SNR fixed at $\text{SNR}_d=4.6~\text{dB}$ and a constant pump power of $1~\text{mW}$. To maintain the same $\text{SNR}_d$ for different $\eta_s^\prime$ values, we only adjust the background noise level ${\mathcal{I}}_s^\prime$. The results from the experiment, as depicted in \figref{fig:Vis_vs_SA}, show that the APV remains nearly constant and closely aligns with our theoretical prediction. It is important to note, however, that with very high signal attenuation, the detected signal count rate is too low for accurate visibility measurement, necessitating long-term data collection for these cases.

\begin{figure}[h!]
\centering
\includegraphics[height=5.5cm]{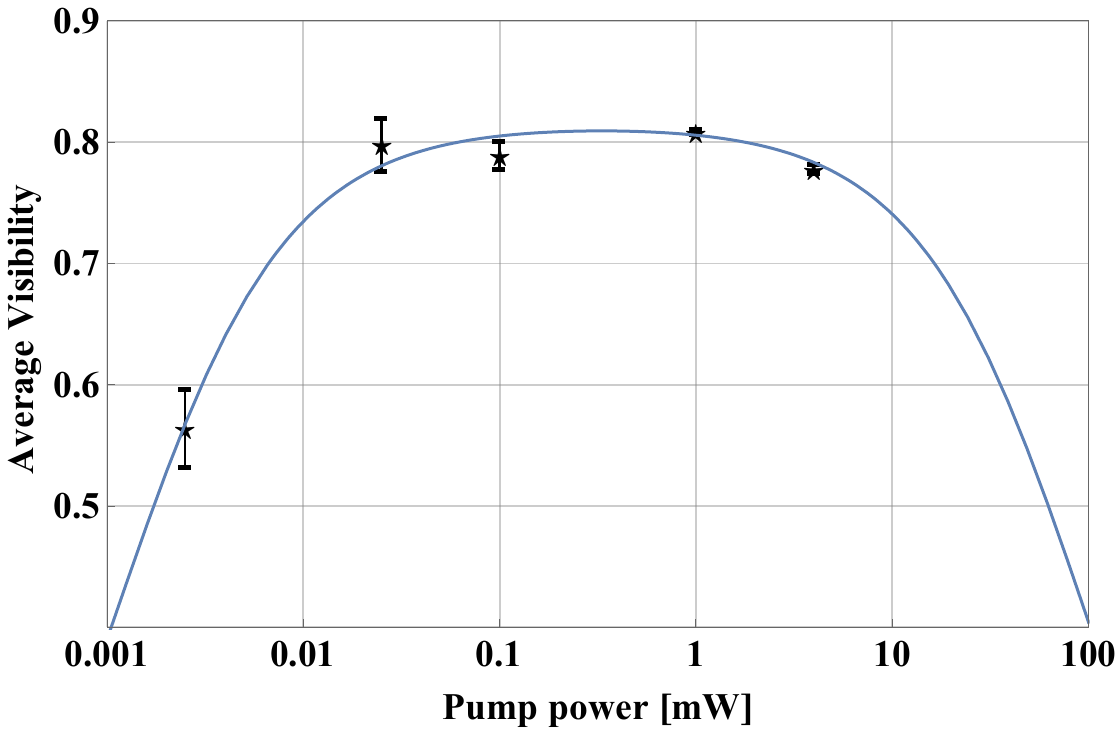}
\caption{Average visibility with varying CW pumping power, for fixed values of uplink channel parameters. $\eta_s^\prime=21~\text{dB}$ attenuation and ${\mathcal{I}}_s^\prime=10^6~\text{cps}$, corresponds to $\text{SNR}_d=6.2~\text{dB}$.}
\label{fig:Vis_vs_pumppower}
\end{figure}

In \figref{fig:Vis_vs_pumppower}, we plot the APV as a function of the brightness of the CW-pumped source, or as a function of pumping power. For this measurement, we fix the $\text{SNR}_d$ value at $6~\text{dB}$, where $\eta_s^\prime=21~\text{dB}$ attenuation and ${\mathcal{I}}_s^\prime=10^6~\text{cps}$. We change the pump power from $2.5~\mu\text{W}$ to $4~\text{mW}$ and measure the APV. For the theoretical plot, we consider the general equations Eq. (\ref{coinc_total}-\ref{eq:bkg-acc}) instead of the simplified \eqref{eq:vis3}, since at lower brightness values $\mathcal{N}$, the background noise and channel efficiency of the idler (${\mathcal{I}}_i^\prime, \eta_i^\prime$) can no longer be neglected. We observe that for a given value of $\text{SNR}_d$, a wide range of optimal pump power values provide maximum achievable APV. Both the extremes of the CW pump power values show a decrease in the APV. High pump power increases the probability of accidental coincidences, whereas low pump power decreases the total signal count rate while the noise count rate remains unchanged.

\begin{figure}[h!]
\centering
\includegraphics[height=5.5cm]{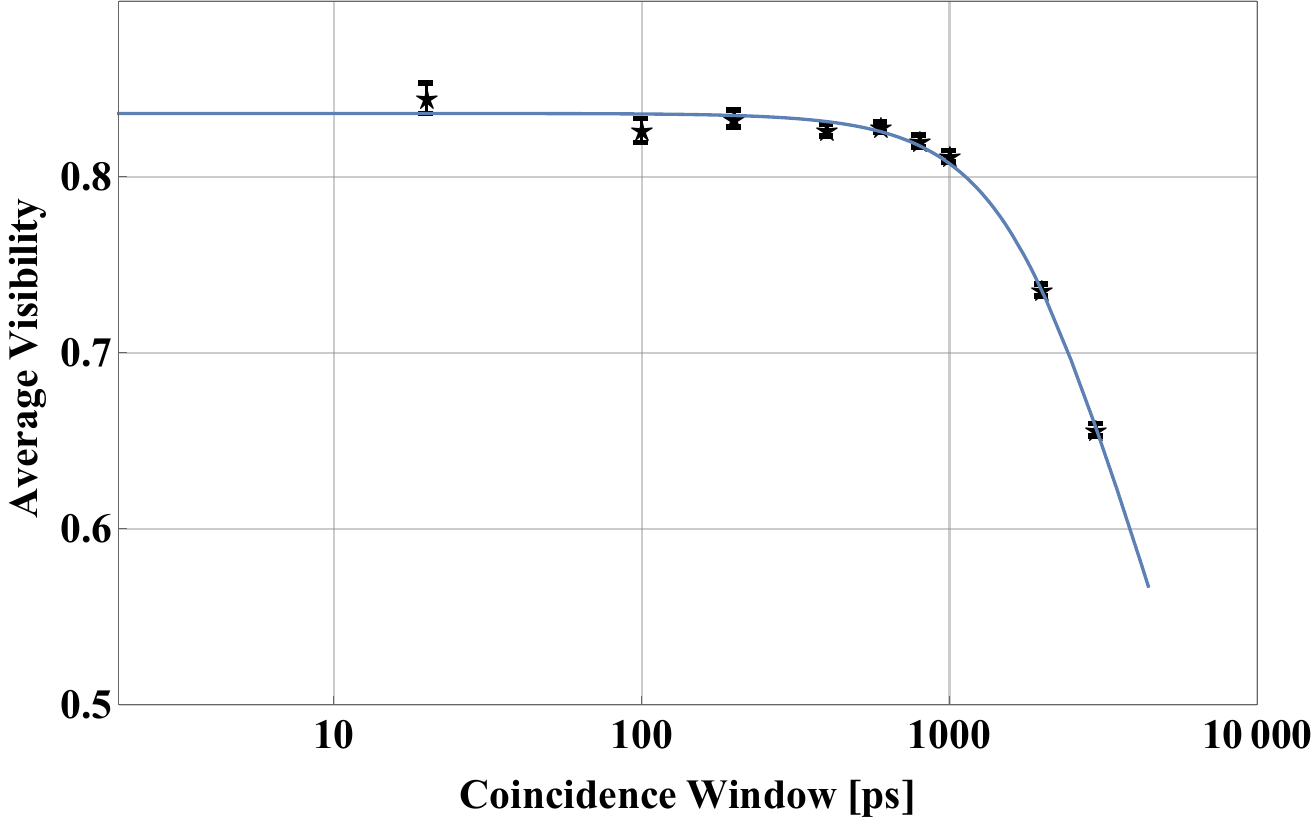}
\caption{Average visibility as a function of the coincidence window, deployed in the post-processing stage. CW pumping power is fixed at $1~\text{mW}$, and $\text{SNR}_d$ value is fixed at $6.2~\text{dB}$.}
\label{fig:Vis_vs_coinwindow}
\end{figure}

We also observe the lowering of the APV with the increasing value of the coincidence timing window in \figref{fig:Vis_vs_coinwindow}. This is consistent with the theoretical model in \eqref{eq:vis3_window} for the case of CW pumped source with a post-selected coincidence window $\tau$. In this plot, we keep the pump power at $1~\text{mW}$, signal attenuation at $\eta_s^\prime=21~\text{dB}$, and background noise at ${\mathcal{I}}_s^\prime=10^6~\text{counts/sec}$. Although the APV saturates at lower $\tau$ values, tight filtering in the timing window also reduces the detected signal photons, leading to a reduction in the SKR in a QKD protocol. Hence, it is important to select an optimal $\tau$ value, which is mainly determined by the timing uncertainty between the arrival times of the signal-idler photons of the same pair. This timing uncertainty is dominated by the timing jitter of the detectors and the accuracy of the time referencing mechanism.

In our final measurement, we perform a proof-of-principal demonstration of entanglement-based QKD. During this demonstration, we extract the raw key rate and QBER from the experimental data. We use a hybrid channel for the QKD, in which the $785~\text{nm}$ signal photons are transmitted through a free-space link simulating an uplink channel of varying length, with additional background noise to replicate daylight operation. The $1651~\text{nm}$ idler photons, on the other hand, are sent through a $10~\text{km}$ long SMF communication channel. 

In the idler arm, the total channel attenuation (${\eta}_i^\prime$), which includes detector and fiber coupling efficiency, is $22~\text{dB}$. The additional detected noise rate (${\mathcal{I}}_i^\prime$) per detector is limited by the dark count rate ($10^3~\text{cps}$). Adding the $10~\text{km}$ SMF introduces an additional delay of approximately $50~\mu\text{s}$ to the idler photons relative to the signal photons. Given that our time tagging device can only delay the electrical signal by a maximum of $100~\text{ns}$, we utilize a delay generator (Stanford Research Systems, DG645) to compensate for this delay. However, it is worth noting that the delay generator's processing capability is limited by the delay applied, restricting the processing count rate to approximately $20~\text{kHz}$ in our case. Therefore, we account for this additional photon loss in the analysis of the total signal channel attenuation.

In \tabref{tab:SecureKeyRate}, we display the raw key rate, QBER, and the asymptotic SKR values for various loss and noise conditions in the uplink channel. SKR is defined here as \cite{neumann2021model}:
\begin{equation}
\label{eq:skr}
   \text{SKR}  = \frac{1}{2} \: \mathcal{R} \:[1 - 2.1 \: \mathcal{H}_{2}(\text{QBER})]
\end{equation}
where $\mathcal{R}$ is the raw key rate which is the summation of all $16$ coincidence count rates, $\mathcal{R}= \sum_{\phi_s,\phi_i\in\mathcal{X}} \sum_{x,y\in\mathcal{Y}}\mathcal{C}_{x,y}^{\phi_s,\phi_i}$. Here, $\mathcal{X} = \{0,\frac{\pi}{4}\}$ is the set of polarization bases and $\mathcal{Y} = \{0,1 \}$ is the set of two orthogonal states for a chosen basis. The factor $\frac{1}{2}$ is due to the basis reconciliation. $\text{QBER}$ is the average of two $\text{QBER}_\phi$, for $\phi\in\{0,\pi/4\}$, defined as $\text{QBER}_\phi=\{{\mathcal{C}}_{0,1}^{\phi}+{\mathcal{C}}_{1,0}^{\phi}\}/\sum_{x,y \in\mathcal{Y}} \mathcal{C}_{x,y}^{\phi}$.

\begin{table}[h!]
\caption{QBER and SKR with asymptotic limit is measured for various combinations of signal channel attenuation and detected background noise.}
\begin{tabular}{|c|c|c|c|c|c|c|}
\hline
\begin{tabular}[c]{@{}c@{}}Pump \\ power\\ {[}mW{]}\end{tabular} & \begin{tabular}[c]{@{}c@{}}Signal \\ attenuation \\ {[}dB{]} \{length\\{[}km{]}\footnote{Signal channel length is estimated from the channel attenuation value, using \eqref{signalorbitcheff}.}\}\end{tabular} & \begin{tabular}[c]{@{}c@{}}Noise\\ (${\mathcal{I}}_s^\prime$)\\ {[}kHz{]}\end{tabular} & \begin{tabular}[c]{@{}c@{}}$\text{SNR}_d$\\ {[}dB{]}\end{tabular} & \begin{tabular}[c]{@{}c@{}}Raw \\ key\\ rate\\ {[}cps{]}\end{tabular} & \begin{tabular}[c]{@{}c@{}}QBER\\ {[}\%{]}\end{tabular} & \begin{tabular}[c]{@{}c@{}}SKR\\ {[}cps{]}\end{tabular} \\ \hline 1 & 25~\{245\} & 0.1 & 42 & 88.3(7)  & 1.7(1)& 32.6(6)\\ 
\hline 4 & 30~\{426\}  & 0.1 & 37 & 119.0(8) & 3.1(1)& 34.5(7)\\ 
\hline 4  & 40~\{1350\}& 2.5& 13 & 15.5(1)& 7.6(2)& 1.4(1) \\ 
\hline 4  & 50~\{4270\}  & 2.5  & 3  & 2.30(5) & 14.52(7)  & NA\footnote{SKR cannot be extracted as the QBER is higher than the threshold value.}\\ \hline
\end{tabular}
\label{tab:SecureKeyRate}
\end{table}
\section{Feasibility of uplink QKD in daylight noise condition}
\label{sec5}
This section demonstrates the feasibility of uplink satellite QKD in a noisy daylight background scenario. Our analysis considers a CW-pumped SPDC source of entangled photon pairs, while all the communication channel parameters are selected based on the current state-of-the-art technical solutions. Our theoretical model, as presented in Section \ref{sec:Experimental Results}, aligns well with the experimental results in a simulated noisy and lossy uplink environment. Therefore, we can confidently extrapolate the model for a realistic implementation of daylight uplink satellite QKD. We investigate the optimal SKR (\eqref{eq:skr}) as a function of various tunable parameters.

To simulate the SKR, we use the formula for coincidences derived from the passive random choice of the measurement basis by the receiver (see \eqref{eqn:qkdmodel2}) in the case of a CW-pumped source, with a coincidence window setup in post-processing (\eqref{eq:vis3_window}). To perform this calculation, we use the reconstructed density matrix obtained from quantum state tomography. Thanks to this approach, we can utilize \eqref{eq:pol_projection1} to compute any required projections for further calculations.

Furthermore, the brightness of the source is $\mathcal{N} = 8 \times 10^6~\text{pairs/sec/mW}$. The uplink channel attenuation is calculated based on \eqref{signalorbitcheff}, with the following parameters: $\eta_{g} = 0.8$, $\eta_{atm} = 0.8$, $R = 15~\text{cm}$, $\lambda = 785~\text{nm}$, $W_{0} = 15~\text{cm}$, and $r_{0} = 8.7~\text{cm}$. $\eta_{r} = 0.3$ is chosen for the case of SMF coupling with FOV of $10~\mu\text{rad}$, following the work in Ref. \cite{liao2017long}. In the case of MMF coupling, we consider a FOV of $100~\mu\text{rad}$ with $\eta_{r} = 0.6$.
The noisy background count rate ${\mathcal{I}}_{s}^\prime$ at the signal receiver, which includes both the dark count rate of the SPAD and the reflected sunlight from the Earth's surface, is calculated using \eqref{eq:SNR_detected} with the following parameters: $a_{E} = 0.3$, $f = 10~\mu\text{rad}$, and $H_{sun} = 4.6 \times 10^{18}~\text{photons/(s-nm-m$^2$)}$. 
The spectral bandwidth of the signal photons is measured to be $0.54~\text{nm}$ FWHM, with the assumption of Gaussian spectral profile, whereas the width of the spectral filter is assumed to be a top hat function and is optimized during the feasibility study. The effect of this filter can be described with an additional efficiency term,
\begin{equation*}
  \eta_{\Delta\lambda}(\sigma_{\lambda,s})=\frac{1}{\sigma_{\lambda,s}\sqrt{2\pi}}\int_{-\Delta\lambda/2}^{\Delta\lambda/2}e^{-(\frac{{\lambda}^2}{2\sigma_{\lambda,s}^2})}d\lambda  
\end{equation*}
where $\sigma_{\lambda,s}$ is the standard deviation of the signal photon spectral profile, and $\Delta\lambda$ is the filter width. This term is included inside the total signal channel efficiency term, $\eta_s^{\prime}\rightarrow\eta_s^{\prime}\times\eta_{\Delta\lambda}(\sigma_{\lambda,s})$. 
The uncertainty in the time difference between the signal-idler wave packets is $\sigma_T=350~\text{ps}$ FWHM, which is mostly contributed by the timing jitter of the detectors. 
The coincidence window ($\tau$) in the feasibility check is chosen in the post-processing stage (see \eqref{eq:vis3_window}) so it can be optimized to find the maximum SKR. 

The total channel attenuation for the idler arm is fixed at $16~\text{dB}$. This accounts for the loss in a $10~\text{km}$ long fiber, as well as fiber-coupling loss and detection efficiency. The background noise rate at the receiver is set to ${\mathcal{I}}_i^{\prime} = 1000~\text{cps}$, consistent with the dark count rate of the SNSPD detectors. 

First, we investigate the optimal pump power for different orbit lengths of the satellite. To do this, we find the maximum SKR for each value of the pump power by using an optimization algorithm while the spectral filter bandwidth ($\Delta\lambda_s$) is varied in the range $1~\text{pm}$ to $2~\text{nm}$ and the coincidence window ($\tau_c$) in the range $1~\text{ps}$ to $3~\text{ns}$. We have chosen $1~\text{pm}$ as the minimum filter bandwidth based on current technical efforts to develop ultra-narrow-band filters \cite{Liu:17,glebov2012volume}. The selection of a $1~\text{ps}$ minimum coincidence window is based on the standard resolution offered by commercial time-tagging devices.

\begin{figure}[h!]
\centering
\includegraphics[height=5.5cm]{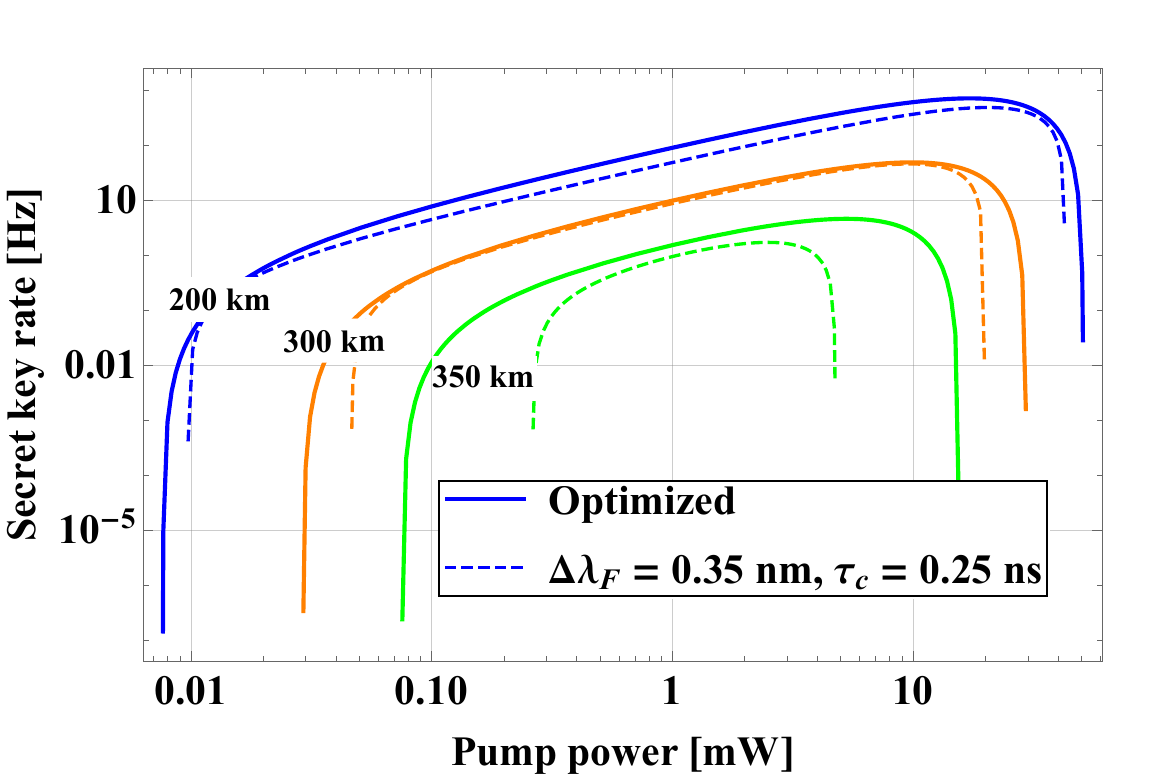}
\caption{SKR as a function of CW pumping power for various lengths of the uplink channel in daylight noise conditions. The solid lines represent optimization in both spectral ($\Delta\lambda_{F}$) and temporal filtering ($\tau_{c}$), while the dashed lines show reduced performance without optimization.}
\label{fig:SKRvsPumppower_opt}
\end{figure}

As shown in \figref{fig:SKRvsPumppower_opt}, we observe that in the low pumping regime, the minimum pump power required to achieve a positive SKR increases with increasing orbit length. This occurs because longer orbit lengths lead to greater attenuation of the signal photon, requiring higher pump power to maintain a similar SNR. Conversely, the maximum pump power that still results in a positive SKR is lower for longer orbits. This is due to the higher probability of accidental coincidence events in the high pumping range for longer orbits, where signal attenuation is more significant. This reduces the APV (alternatively, increases the QBER) and restricts the pumping range at higher pump power values.

Additionally, the optimal pump power changes depending on the length of the orbit. For instance, if the orbit length is $350~\text{km}$, a pump power of $7~\text{mW}$ can achieve an SKR of approximately $5~\text{Hz}$. This observation is significant because it shows that the optimal pump power can be adapted based on the changes in the channel attenuation. Consequently, a feedback loop can dynamically re-adjust the pump power to accommodate weather conditions and variations in the satellite's position. This is especially advantageous in an uplink scenario, where the source is located in the ground station, allowing for a more complex setup without the constraints of space-qualified components.
We also plot the non-optimized behavior with a dashed line. In this case, we use a fixed spectral filter width of $0.35~\text{nm}$ and a coincidence window of $0.25~\text{ns}$. Without spectro-temporal optimization, the potential pumping regions are limited, leading to a lower maximum achievable SKR. 

\begin{figure}[h!]
\centering
\includegraphics[height=5.5cm]{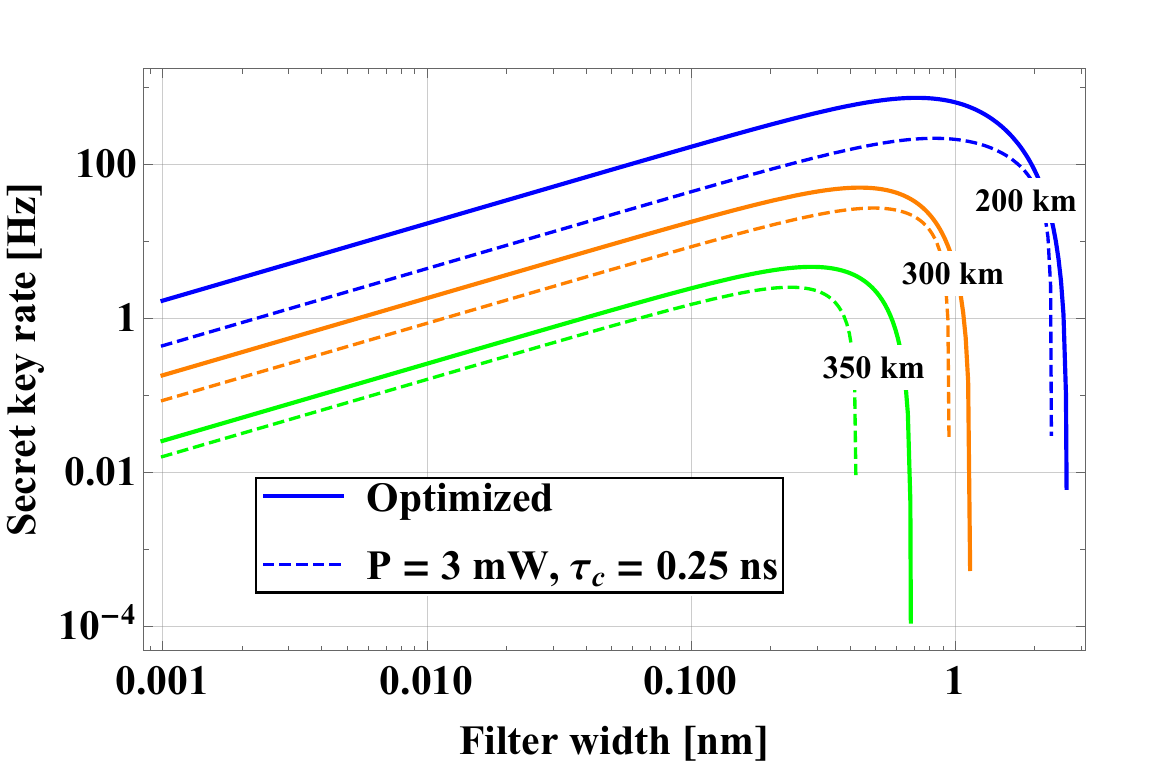}
\caption{SKR as a function of spectral filter width for different uplink channel lengths in daylight noise conditions. Solid lines represent optimization in CW pump power ($P$) and temporal filtering ($\tau_{c}$), while dashed lines indicate no optimization.}
\label{fig:FilterwidthvsSKR_opt}
\end{figure}

In \figref{fig:FilterwidthvsSKR_opt}, we examine the best spectral filter bandwidth in relation to the SKR, for various orbit lengths. To achieve this, each data point (solid line) is optimized for pump power ranging from $1~\mu\text{W}$ to $1~\text{W}$ and the coincidence window spanning from $1~\text{ps}$ to $3~\text{ns}$.
We notice that if the filter width is too narrow, the SKR decreases because the signal photons are filtered out more compared to the noisy photons. 
On the other hand, if the spectral filter is too wide for a fixed orbital length, the SKR quickly drops to zero. This happens because too many noisy photons are captured by the receiver compared to the signal photon, reducing the SNR to a level that is insufficient to extract any positive secret key. This behavior also changes with the length of the orbit. The channel attenuation increases for greater lengths, resulting in fewer received photons. More stray photons must be filtered out to maintain sufficient SNR for secret key extraction. However, achieving any arbitrary orbit length is impossible because excessively tight filtering for longer orbits leads to a significant loss of signal photons compared to stray photons.
Furthermore, we demonstrate that for a fixed value of the coincidence window and pump power, the range of operating filter width, as well as the maximum achievable SKR is reduced (dashed line in \figref{fig:FilterwidthvsSKR_opt}).

\begin{figure}[h!]
\includegraphics[height=5.5cm]{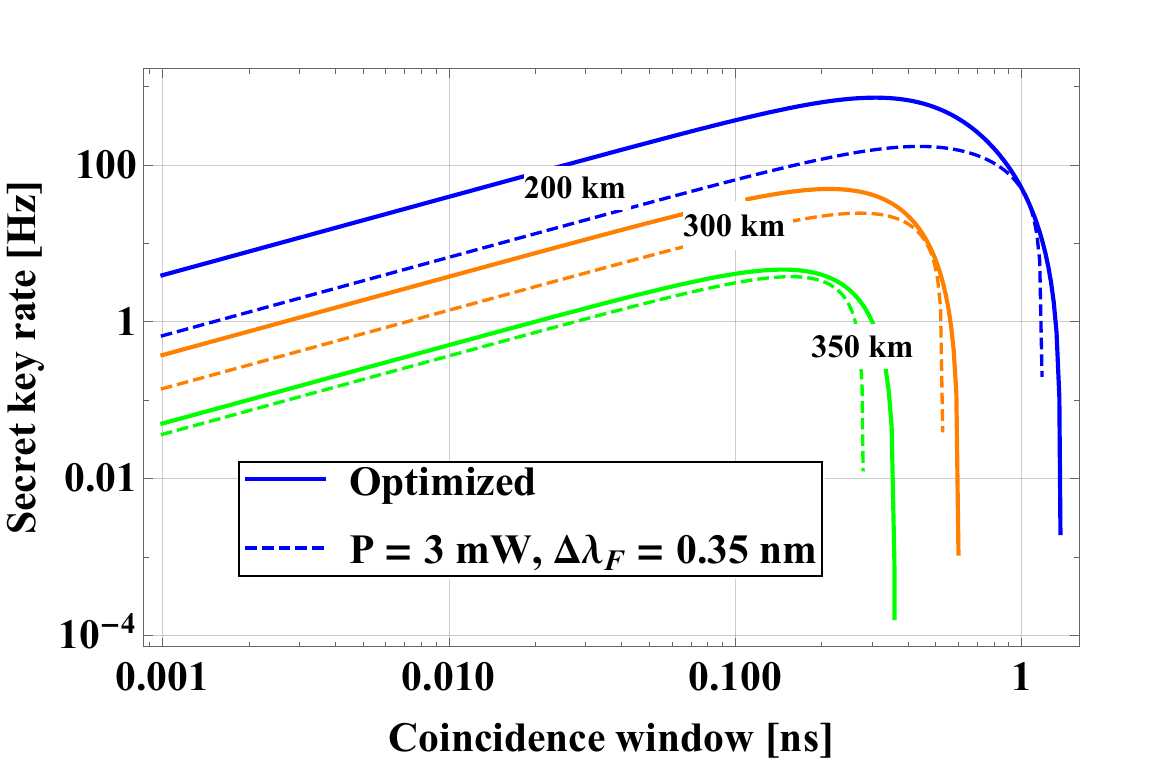}
\caption{SKR as a function of coincidence timing window for different uplink channel lengths in daylight noise. Solid lines represent optimized CW pump power ($P$) and spectral filtering ($\Delta\lambda_{F}$), while dashed lines indicate no optimization.}
\label{fig:CoinFiltrerVSSKR_opt}
\end{figure}

In \figref{fig:CoinFiltrerVSSKR_opt}, we explore the optimization of temporal filtering by adjusting the coincidence window during the post-processing stage, similar to the optimization of spectral filtering. We observe a behavior comparable to the case of spectral filtering as shown in \figref{fig:FilterwidthvsSKR_opt}. If the coincidence window is too small, it predominantly reduces the detection of signal photons, decreasing the SKR. Conversely, if the coincidence window is too large, it results in too many accidental coincidences, making it impossible to extract the secret key. Each point on the solid line in \figref{fig:CoinFiltrerVSSKR_opt} undergoes optimization for a pump power range of $1~\mu\text{W}$ to $1~\text{W}$ and spectral filter width varying from $1~\text{pm}$ to $2~\text{nm}$. Additionally, the dashed line represents a fixed value of pump power and spectral filter width for better comparison with the effect of the optimization process.

\begin{figure}[h!]
\includegraphics[height=5.5cm]{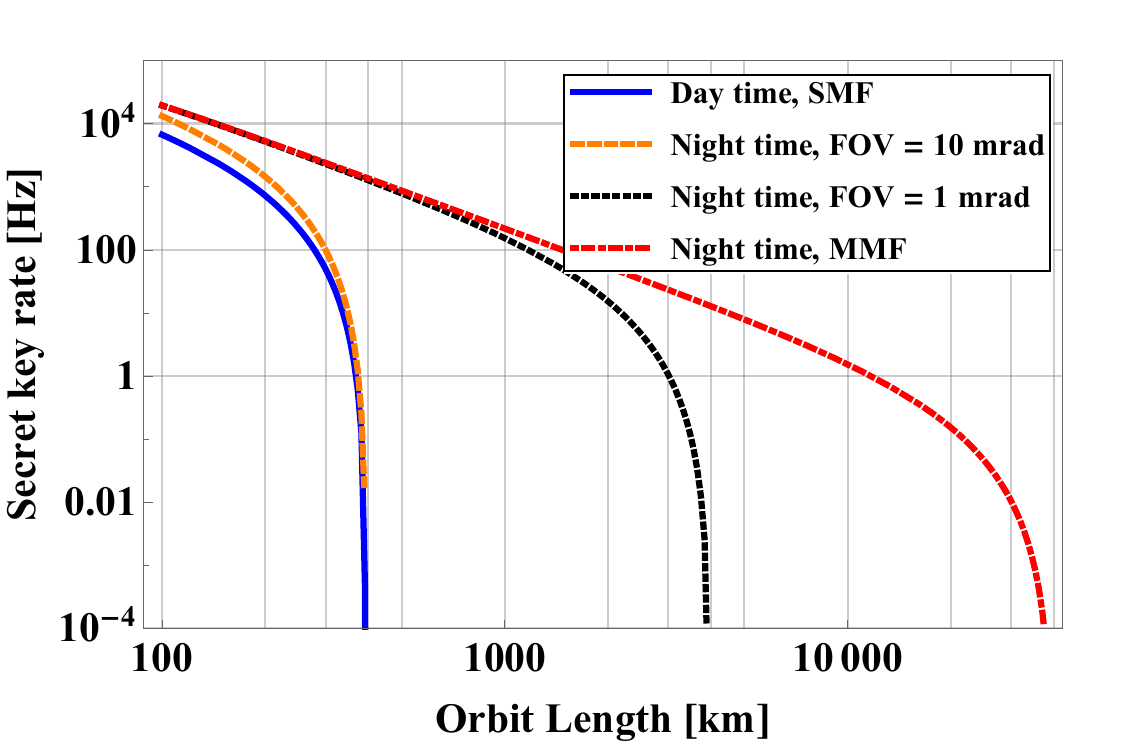}
\caption{Maximum achievable length of the uplink satellite capable of extracting a positive SKR. Both daytime and nighttime noise conditions are considered with various FOVs of the receiver satellite. }
\label{fig:ORbit_lentgh}
\end{figure}

We also calculate the longest distance that can be achieved using the state-of-the-art technical solutions available for entangled photon source and uplink satellite communication. In \figref{fig:ORbit_lentgh}, we present the simulation results for both daytime and nighttime background conditions, taking into account SMF and MMF coupling with different FOVs of the receiver telescope.
The results indicate that by utilizing SMF coupling in combination with optimal filtering and pumping conditions, it is possible to achieve an orbit length of $ 400~\text{km}$, which is suitable for lower-range LEO satellites, even in the presence of daylight background noise. However, with standard MMF coupling (FOV $>100~\mu\text{rad}$), QKD in an uplink-daylight condition is not feasible, even with the best available sources of entangled photons based on SPDC. In fact, the maximum FOV that can provide a positive SKR in LEO orbits is close to $27~\mu\text{rad}$. This limitation could be overcome by further reducing the spectral bandwidth of the entangled photons and minimizing the overall timing jitter of the QKD system.

When reducing the FOV of the receiver telescope, it is important to consider the decrease in photon collection efficiency, particularly when using SMF coupling. In this situation, incorporating adaptive optics \cite{pugh2020adaptive,acosta2024analysis} into the receiver satellite is essential for achieving optimal mode matching.

Furthermore, we test the feasibility of the uplink QKD in a nighttime noise condition when the stray photons are significantly lower compared to the daytime condition. Our results demonstrate that with a FOV of $10~\text{mrad}$, we can achieve QKD over a distance of $400~\text{km}$, while a FOV of $1~\text{mrad}$ allows for positive SKR generation up to a distance of $4000~\text{km}$. Notably, when the receiver telescope utilizes MMF coupling with a FOV of $100~\mu\text{rad}$, it becomes feasible to perform QKD at the geostationary orbits (GEO) satellite distance of approximately $40 000~\text{km}$.

\begin{figure}[h!]
\includegraphics[height=5.5cm]{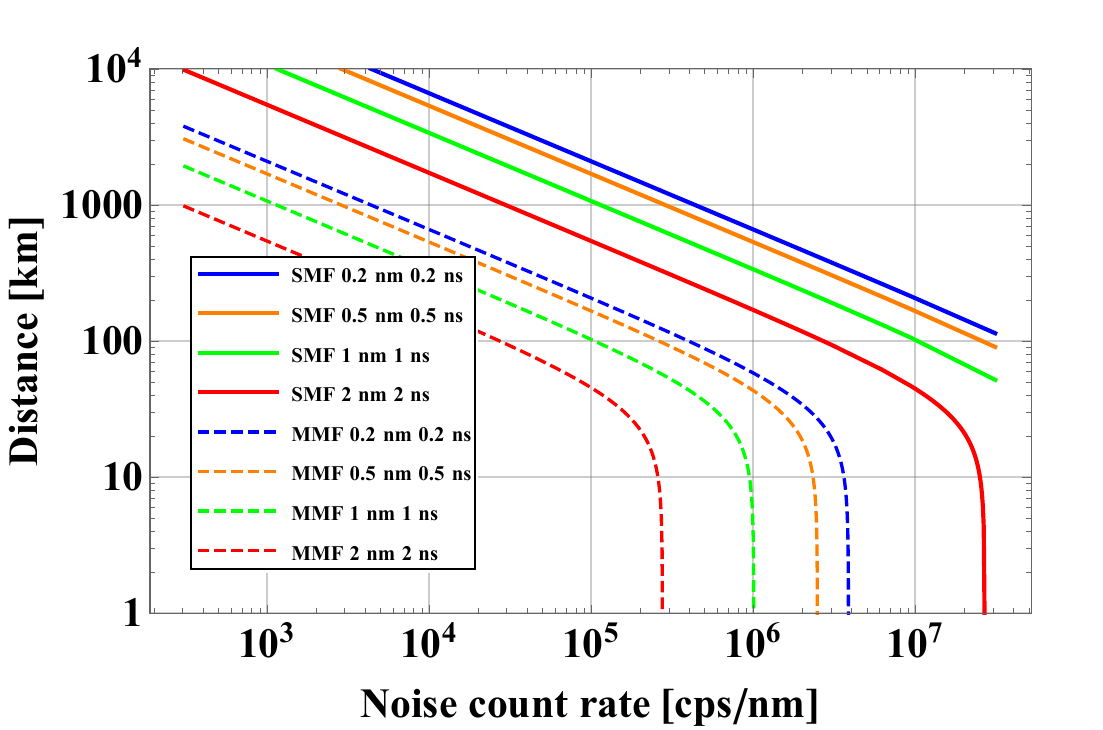}
\caption{Maximum achievable distance of the uplink satellite as a function of the noise photon count rate collected in an SMF or MMF at the receiver. The effect of tight spectro-temporal filtering is also illustrated.}
\label{fig:distance_iday}
\end{figure}

In figure \ref{fig:distance_iday}, we illustrate the maximum achievable distance for a positive SKR as it varies with different noise levels, coupling conditions, and filtering conditions. We consider both SMF and MMF coupling for various combinations of spectral filter width and coincidence timing window. The noise count rate on the x-axis refers to the number of noisy photons per unit time at the receiver telescope, which is coupled to an SMF or MMF. Under daytime conditions, a typical noise count rate for a $15~\text{cm}$ radius receiver telescope is approximately $3.1 \times 10^{6} ~\text{cps/nm}$ when using an SMF, and $3.1 \times 10^{8} ~\text{cps/nm}$ when using an MMF. At night, this noise count rate is drastically reduced by an order of six.

\figref{fig:distance_iday} indicates that operating under daylight conditions at an LEO orbit distance is not feasible in the MMF scenario. When employing MMF coupling, the operational range relies on the choice of filters. However, the studied scenarios suggest that operating at LEO orbit commences at a noise level of approximately $10^{4}~\text{cps/nm}$. On the other hand, SMF coupling allows operation in daylight conditions, given that the spectral and temporal filters are stricter than $1~\text{nm}$ and $1~\text{ns}$, respectively. It is possible to achieve this with SMF coupling due to its significantly narrower FOV than MMF coupling.

Furthermore, in all cases, we observe that as the noise count rate decreases, the maximum achievable distance increases. This happens because a higher attenuation allows for the required SNR for the secure key exchange to be maintained over a longer distance due to fewer noise photons. On the contrary, when the noise count rate is too high, the maximum distance sharply decreases, approaching zero. We also show that stricter filtering increases the tolerance for higher noise count rates. 

\section{Discussion and Outlook}
\label{sec6}
This work shows the feasibility of the uplink-daylight QKD using a realistic polarization-entangled photon-pair source based on the SPDC process. Our analysis concludes that, with the current state-of-the-art technology and SMF coupling, we can perform secure QKD in a daylight-uplink condition up to an LEO satellite distance of $400~\text{km}$. However, with standard MMF coupling daylight QKD is not feasible for this range. MMF coupling can only reach distances suitable for even GEO satellites during nighttime operations.

Also, we show the significant enhancement of SKR by only optimizing the pumping power of the source, filter bandwidth in the receiver telescope, and temporal coincidence window. This conclusion suggests that one may use an active feedback loop to tune these three parameters such that the SKR value is tolerant against any local atmospheric fluctuations (cloud, storm, rain, etc.) and changes in satellite elevation angle. Our proposed theoretical model is equally suitable for both the CW and pulse-pumped SPDC sources and is consistent with the experimental results. 

We have successfully demonstrated the experimental distribution of entanglement through a noisy uplink channel, mimicking both daytime and nighttime conditions. These simulations were performed in a controlled laboratory environment to ensure accurate replication of real-world conditions. The results demonstrated the robustness of the entanglement distribution process in various background noise levels and channel losses, confirming the adaptability and effectiveness of the model in different atmospheric conditions.

Furthermore, we presented a QKD demonstration employing non-degenerate (dual-wavelength) entangled photons using a hybrid communication channel that combines free-space and fiber optics. This hybrid setup effectively simulated real-world application scenarios. We achieved an SKR of $1.4~\text{cps}$, with a QBER of $7.6~\text{\%}$. The free-space channel experienced a loss of $40~\text{dB}$, which is equivalent to an uplink distance of approximately $1350 ~\text{km}$, while the fiber channel extended for a distance of $10 ~\text{km}$.

To improve the performance of QKD in the uplink scenario and under daylight conditions, one may further optimize the entangled photon source. 
One possible approach is to increase the brightness of the source by utilizing a longer non-linear crystal for SPDC. A longer crystal is expected to result in higher brightness per unit pump power, showing an approximately linear dependence \cite{fedrizzi2007wavelength}.
Another step is to prepare a narrower photon spectrum by using a cavity-based SPDC source \cite{slattery2019background}. This approach can produce spectral linewidth on the order of MHz, allowing for very narrow spectral filtering at the receiver. However, sources at this stage do not have the same brightness, so their effects need to be studied further.

Furthermore, switching from a CW-pumped source to a pulse-pumped source offers several advantages. One key benefit is that the detector at the receiver will only need to be active during the expected arrival of the signal. This selective activation can significantly improve the SNR by reducing the likelihood of detecting accidental coincidences, particularly in daylight or high-noise environments. Additionally, a pulsed laser with a high repetition rate can generate more photon pairs, while the multi-photon generation probability remains unchanged. This enhances the SNR further by increasing the number of signal photons detected and overall SKR. The combination of precise timing control, reduced noise interference, and higher photon generation makes pulsed lasers an attractive option for quantum communication, particularly in satellite-based QKD systems, where time-synchronization and efficient use of detection resources are critical.

Moreover, efforts can be made to minimize the system's timing jitter, allowing for a much smaller coincidence window. This jitter reduction improves the SNR by enabling more precise temporal filtering, which helps block out noisy or stray photons outside the expected time window for photon detection.

\section*{acknowledgments}

The authors thank Thomas Jennewein for insightful discussions and acknowledge financial support from Horizon Europe, the European Union’s Framework Programme for Research and Innovation, SEQUOIA project, under grant agreement 101070062.

\bibliography{sample}

\end{document}